\documentclass[aps,prb,reprint,nofootinbib,twocolumn,superscriptaddress,showpacs,showkeys,longbibliography,amsmath,amssymb]{revtex4-1}
\usepackage{graphicx}% Include figure files
\usepackage{dcolumn}% Align table columns on decimal point
\usepackage{bm}% bold math
\usepackage{braket}
\usepackage{subfigure}
\usepackage[colorlinks,bookmarks=false,citecolor=blue,linkcolor=red,urlcolor=blue]{hyperref}
\usepackage[english]{babel}

\begin{document}
\preprint{APS/123-QED}

\title{Dynamics of a quantum phase transition in the Aubry-Andr\'{e}-Harper model with $p$-wave superconductivity}

\author{Xianqi Tong}
\affiliation{Department of Physics, Zhejiang Normal University, Jinhua 321004, People's Republic of China}
\author{Yeming Meng}
\affiliation{Department of Physics, Zhejiang Normal University, Jinhua 321004, People's Republic of China}
\author{Xunda Jiang}
\affiliation{Guangdong Provincial Key Laboratory of Quantum Metrology and Sensing \& School of Physics and Astronomy, Sun Yat-Sen University (Zhuhai Campus), Zhuhai 519082, China}
\author{Chaohong Lee}
\affiliation{Guangdong Provincial Key Laboratory of Quantum Metrology and Sensing \& School of Physics and Astronomy, Sun Yat-Sen University (Zhuhai Campus), Zhuhai 519082, China}
\affiliation{State Key Laboratory of Optoelectronic Materials and Technologies,
Sun Yat-Sen University (Guangzhou Campus), Guangzhou 510275, China}
\author{Gentil Dias de Moraes Neto}
\affiliation{Department of Physics, Zhejiang Normal University, Jinhua 321004, People's Republic of China}
\author{Gao Xianlong}
\affiliation{Department of Physics, Zhejiang Normal University, Jinhua 321004, People's Republic of China}

\date{\today}% It is always \today, today, 	
\begin{abstract}
We investigate the nonequilibrium dynamics of the one-dimension Aubry-Andr\'{e}-Harper model with $p$-wave superconductivity by changing the potential strength with slow and sudden quench. Firstly, we study the slow quench dynamics from localized phase to critical phase by linearly decreasing the potential strength $V$.  The localization length is finite and its scaling obeys the Kibble-Zurek mechanism. The results show that the second-order phase transition line shares the same critical exponent $z\nu$, giving the correlation length $\nu=0.997$ and dynamical exponent $z=1.373$, which are different from the Aubry-Andr\'{e} model. Secondly, we also study the sudden quench dynamics between three different phases: localized phase, critical phase, and extended phase. In the limit of $V=0$ and $V=\infty$, we analytically study the sudden quench dynamics via the Loschmidt echo. The results suggest that, if the initial state and the post-quench Hamiltonian are in different phases, the Loschmidt echo vanishes at some time intervals. Furthermore, we found that, if the initial value is in the critical phase, the direction of the quench is the same as one of the two limits mentioned before, and similar behaviors will occur.
\end{abstract}

\pacs{Valid PACS appear here}% PACS, the Physics and Astronomy
\maketitle

%\tableofcontents

\section{Introduction}
\label{intro}
In recent years, extensive researches have
been carried to unravel the behavior of quasiperiodic (QP) structures\cite{QP1,QP2,QP3,QP4,QP5,QP6}. QP system, being aperiodic but deterministic, lacks translational invariance but shows long-range order leading to a rich critical behavior. The critical properties are different or can be regarded as intermediate from those of ordinary (periodic) and disordered (random) systems. For instance, the spatial modulation of the parameters can change the universality class of a quantum phase transition (QPT), i.e.  the critical exponents that  characterized the equilibrium properties of the physical observables at the transition point. Furthermore, one-dimensional (1D) QP systems, known as the Aubry-Andr\'{e}-Harper (AAH) model, show Anderson localization transition at a finite strength of the QP disorder that differs from the original 1D random model.

In the AAH model, the states at the critical point are neither extended nor localized but critical, characterized by power-law localization, and fractal-like spectrum and wave functions. In an interacting system, the many-body localization with random or QP case exhibit quite different behaviors\cite{SP}.
Furthermore, the quantum phase transitions of QP system related to quantum magnetism described by spin Hamiltonians \cite{SP1,SP2,SP3,SP4,SP5,SP6,SP7,SP8,SP9,SP10} and respective fermionic counterpart\cite{fermi1,fermi2,fermi3}, were studied extensively. In particular, the anisotropic XY chain in a transverse magnetic field\cite{XY1,XY2,XY3,XY4,XY5,XY6,XY7}, that maps via Jordan-Wigner transformation, to  the AAH model with $p$-wave superconducting (SC) pairing terms\cite{fermi1,fermi2,fermi3}, and contains the quantum Ising and XY chains as limiting cases, has drawn attention for a rich phase diagram, as depicted in Fig.~\ref{phase diagram}. The anisotropy (SC pairing) destroys the self-duality of the isotropic XY model and stabilizes the critical phase sandwiched between extended and localized phases.

Although, the phase diagram of the AAH model with SC pairing is well understood, it lacks the thorough investigation of the critical behavior and the nonequilibrium dynamics. Specifically, in a continuous phase transition, the correlation length $\xi$ and corresponding gap $\Delta$ diverge at the transition as $\xi\approx\epsilon^{-\nu}$ and $\Delta \approx\xi^{-z}$, where $\epsilon$ is the distance from the critical point of the disorder strength, $\nu$ and $z$ are the correlation length and dynamical critical exponents. To the best of our knowledge, there is no report in literature about the critical exponents of the  AAH model with SC pairing.

In this context, it is important to determine the nonequilibrium dynamical signatures of a quantum phase transition (QPT)\cite{RefDA,RefDB,RefDM,RefDC,RefDF,RefDG,RefDL,RefDN,RefDQ,RefDR,RefDV,RefDI,RefDT,RefDT1,RefDT2}, which has also being explored in QP system both experimentally\cite{RefBI,RefBJ,RefBK} and theoretically\cite{SP1,SP2,SP3,SP4,RefBH,RefBH1,RefBH2,RefBH3,RefBH4}. It is useful to discriminate between two limiting processes of slowly and instantaneously changing of the parameters. Driving the parameter across the second-order phase transition is usually described by the Kibble Zurek mechanism (KZM)\cite{RefBB,RefBB1,RefBB2}. The essence of the KZM is the breaking of the adiabaticity for crossing the critical point of a QPT, which leads to the corresponding excitations following a power law relation with respect to the quench rate. For the dynamical quantum phase transition (DQPT), the quantum system is quenched out of equilibrium by suddenly changing the parameters of the Hamiltonian. For the sudden quench dynamics, Loschmidt echo is an important quantity, which measures the overlap between the initial state and the time-evolved state\cite{RefDL,RefDN,RefDQ,RefDR}. Many theoretical works have demonstrated that the Loschmidt echo plays a significant role in characterizing the nonequilibrium dynamical signature of the quantum phase transition\cite{RefDA,RefDB,RefDM,RefDN}. Recently, thanks to the developments of the quantum simulation techniques, DQPT can be directly detected in a string of ions simulating the interacting transverse field Ising model\cite{RefDP}.

However, the time evolution of the Loschmidt echo and the KZM requires an in-depth investigation in a 1D QP system which exists phase transitions among localized phase, critical phase, and extended phase. Here, we pay attention to such a  quantum disordered system described by the AAH model with $p$-wave SC paring \cite{RefJE,RefB,RefJ,RefAU,RefJF,RefJG}.

The rest of the paper is organized as follows. In Sect.~\ref{sec:2}, we explicitly write down the Schr\"{o}dinger equation of the 1D QP system. In Sect.~\ref{sec:3}, we calculate the critical exponents and verify the KZM hypothesis. In Sect.~\ref{sec:4}, we discuss the sudden quench dynamics of the quantum phase transition between different phases and give the analytical expressions of two limits cases. Section~\ref{sec:5} is devoted to conclusion.

\section{Model Hamiltonian}
\label{sec:2}
\begin{figure}[h]
\centering
\includegraphics[width=0.5\textwidth]{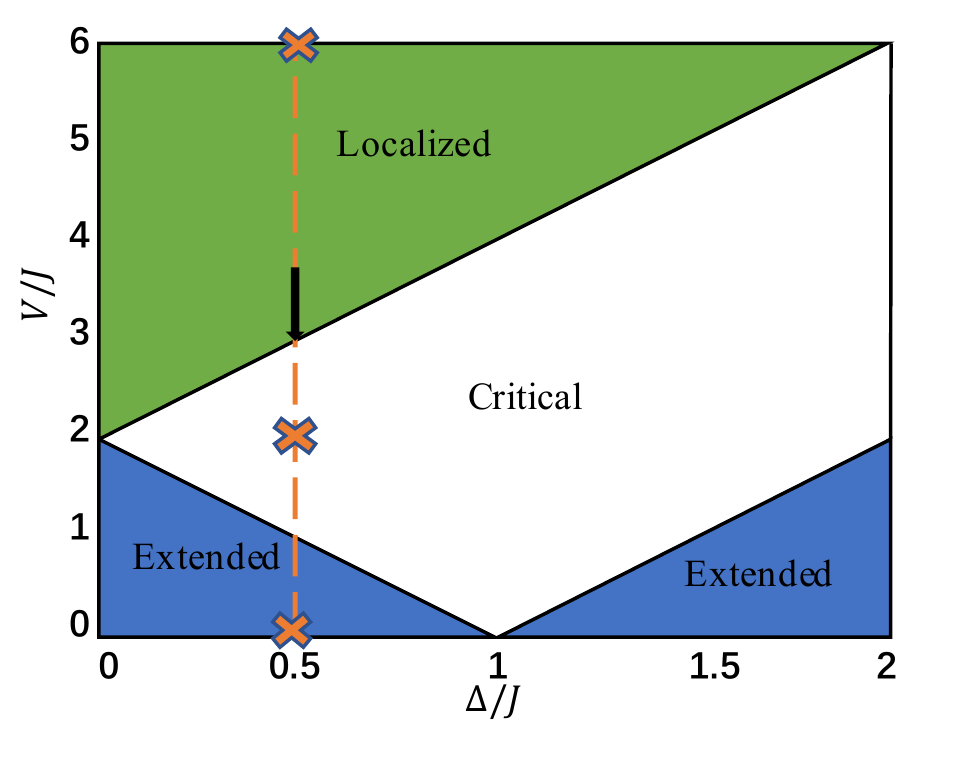}
%\caption{fig1}
\caption{(Color online) Sketch of the phase diagram of the AAH model with $p$-wave superconducting paring order parameter $\Delta$ and the disorder strength $V$\cite{RefB}. Three different phases, that is, extended phase, critical phase and localized phase are shown up in different parameter regimes. The line between critical phase and localized phase is a second-order phase transition line. The vertical line at $\Delta=0.5$ passing through three different phases indicates the cases we studied. The KZM as indicated by the solid arrow is studied in Fig.~\ref{power law.a}, and the quenching as indicated by the crosses is studied in Figs.~\ref{Los-Ex} and ~\ref{Los-lc}.}
\label{phase diagram}
\end{figure}

The generalized 1D AAH model with $p$-wave SC paring is described by the following Hamiltonian
\begin{equation}
\label{E1}
H=\sum_{j=1}^{N}V_{j}c_{j}^{\dagger}c_{j}+\sum_{j=1}^{N-1}(-Jc_{j+1}^{\dagger}c_{j}+\Delta c_{j+1}^{\dagger}c_{j}^{\dagger}+H.c.),
\end{equation}
where $c_{j}$ $(c_{j}^{\dagger})$ is the fermionic annihilation (creation) operator at the $j$-th site. Here $V_{j}=V\cos(2\pi\alpha j+\phi)$ is the incommensurate potential with $\alpha$ = $(\sqrt{5}-1)/2$ being an irrational number, $V$ is the strength of the incommensurate potential, and the random phase $\phi \in[0,2\pi)$ is introduced as a pseudorandom potential. $J$ is the nearest-neighbor hopping amplitude and we set $J=1$ as energy unit throughout this paper. $\Delta$ is the amplitude of the $p$-wave SC paring. The phase diagram of this system has three different phases shown in Fig.~\ref{phase diagram}: localized phase, critical phase and extended phase, which are marked by green, white and blue, respectively. For $V=2|J+\Delta|$, the system undergoes a second-order phase transition from critical phase to localized phase\cite{RefJE}. For $V=2|J-\Delta|$, the system has a phase transition from critical phase to extended phase\cite{RefB,RefJ}. Firstly, we need to rewrite the Hamiltonian by using the Bogoliubov-de Gennes (BdG) transformation,\\
\begin{equation}
\eta^{\dagger}_{n}=\sum_{j=1}^{N}[{u_{n,j}c_{j}^{\dagger}+v_{n,j}c_{j}}],
\end{equation}
where $n=1,...,N,$ the Bogoliubov modes $(u_{n,j},v_{n,j})$ are the eigenstates of the Hamiltonian and $u_{n,j},v_{n,j}$ are chosen be real, so the Hamiltonian can be diagonalized as \\
\begin{equation}
H=\sum_{n=1}^{N}{\varepsilon_{n}(\eta_{n}^{\dagger}\eta_{n}}-\frac{1}{2}),
\label{diag}
\end{equation}
with $\varepsilon_{n}$ being the spectrum of quasiparticles. For the $n$-th Bogoliubov modes, we have the following BdG equations:

\begin{equation}
\begin{array}{l}
-J u_{ j-1}+\Delta v_{ j-1}+V_{j} u_{ j}-J u_{ j+1}-\Delta v_{ j+1}=\varepsilon u_{ j}, \\
-\Delta u_{ j-1}+J v_{ j-1}-V_{j} v_{ j}+\Delta u_{ j+1}+J v_{ j+1}=\varepsilon v_{ j}.
\end{array}
\end{equation}
The wave function is expressed as
\begin{equation}
\ket{\Psi_{n}}=[u_{n,1},v_{n,1},u_{n,2},v_{n,2},...,u_{n,N},v_{n,N}]^{T},
\end{equation}
then for the Schr\"{o}dinger equation $H\ket{\Psi_{n}}=\varepsilon_{n}\ket{\Psi_{n}}$, the Hamiltonian can be written as a $2N\times2N$ matrix:
\begin{equation}
H=\left(\begin{array}{ccccccc}
A_{1} & B & 0 & \ldots & \ldots & \ldots & C\\
B^{\dagger} & A_{2} & B & 0 & \ldots & \ldots & 0\\
0 & B^{\dagger} & A_{3} & B & 0 & \ldots & 0\\
\vdots & \ddots & \ddots & \ddots & \ddots & \ddots & \vdots\\
0 & \ldots & 0 & B^{\dagger} & A_{N-2} & B & 0\\
0 & \ldots & \ldots & 0 & B^{\dagger} & A_{N-1} & B\\
C^{\dagger} & \ldots & \ldots & \ldots & 0 & B^{\dagger} & A_{N}
\end{array}\right),
\end{equation}
where\\
\begin{equation}
A=\left(\begin{array}{cc}
V_{j} & 0\\
0 & -V_{j}
\end{array}\right),
\end{equation}
\begin{equation}
B=\left(\begin{array}{cc}
-J & -\Delta\\
\Delta & J
\end{array}\right),
\end{equation}
and
\begin{equation}
C=\left(\begin{array}{cc}
-J & \Delta\\
-\Delta & J
\end{array}\right).
\end{equation}
Here, we assume the Hamiltonian with periodic boundary condition, hence $\alpha$ can be approximated by a rational number with $L$ in the denominator. Dependence of $L$ implies an order $L=F_{m}, \alpha=F_{m-1}/F_{m}$, where $F_{m}$ is a Fibonacci number.\\

\begin{figure}
\begin{centering}
\includegraphics[width=1.66in]{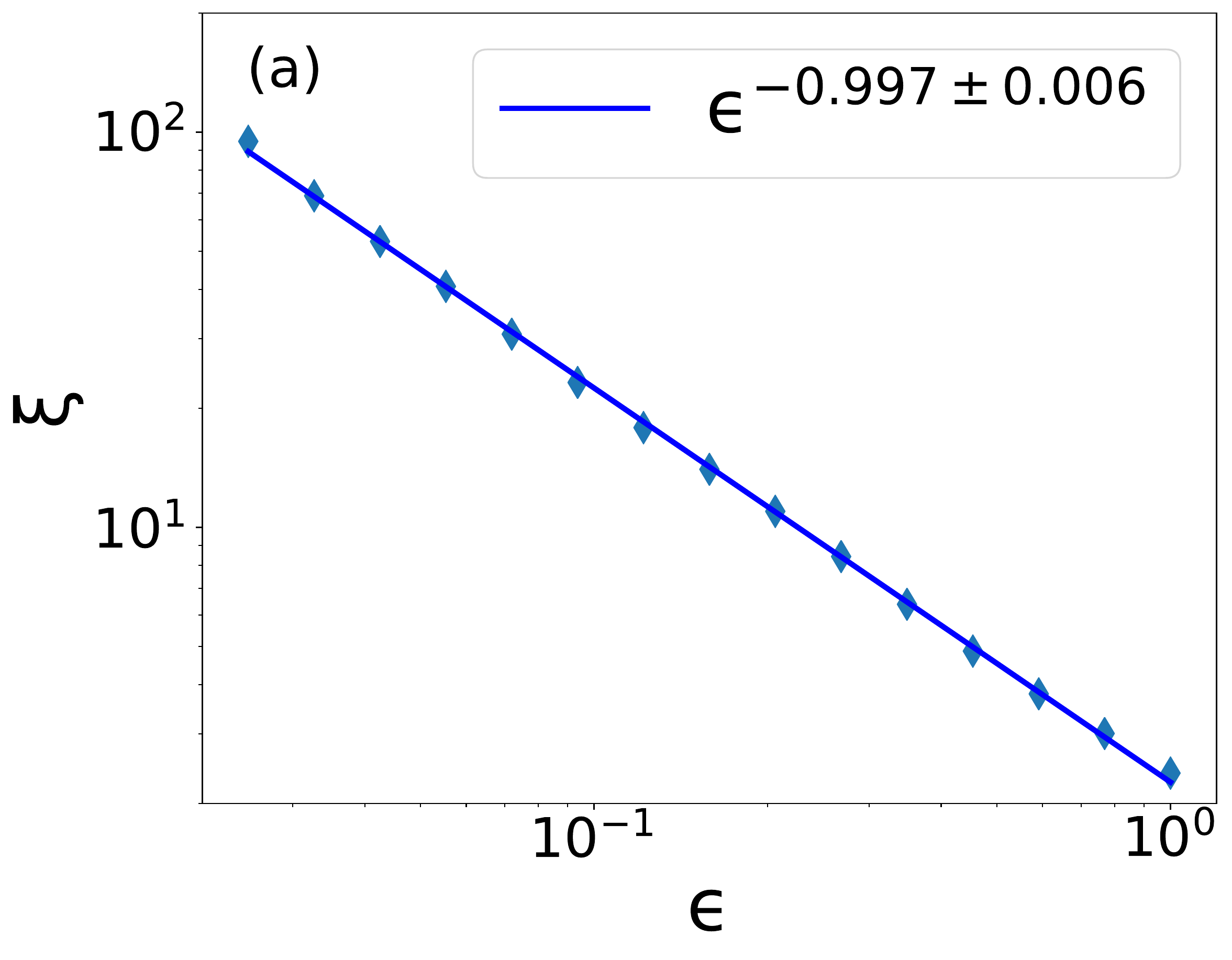}\includegraphics[width=1.7in]{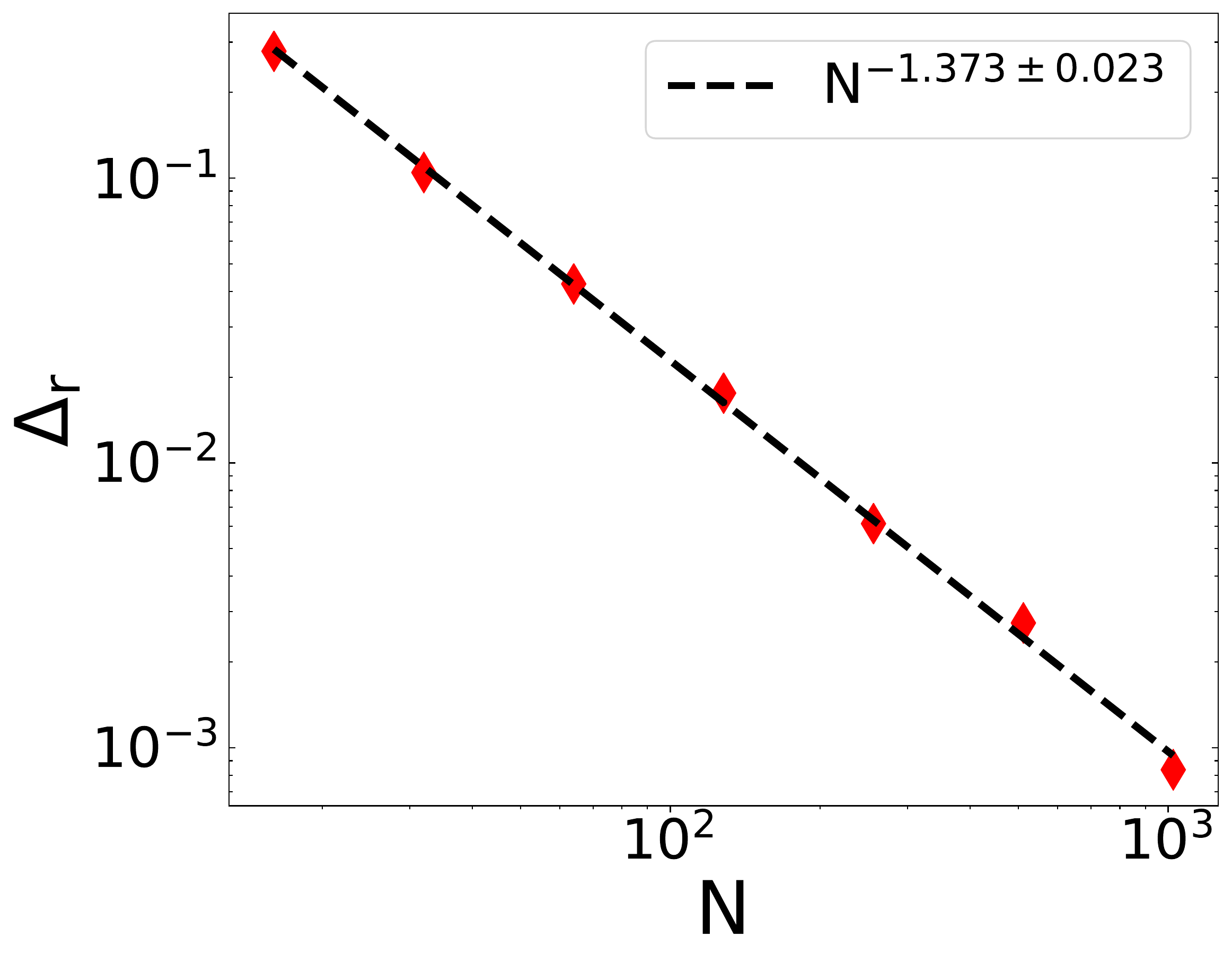}

\caption{(a) The localization length $\xi$ as a function of the distance from the critical point $\epsilon=V-V_{c}$. Here $\xi$ was calculated by using the ground state of the corresponding Hamiltonian. The linear fit $\xi\sim\epsilon^{-\nu}$ yields correlation-length exponent $\nu=0.997\pm0.006$; cf.~Eq.~(\ref{scaling_length}). (b) the relevant gap $\Delta_{r}$ which is the sum of two lowest eigenenergies at the critical point as a function of $N$. Fitting $\Delta_{r}\sim N^{-z}$ yields a dynamical exponent $z=1.373\pm0.023$; see Eq.~(\ref{scaling_gap}). Here, we set SC paring amplitude $\Delta$=0.5, the critical point $V_{c}$=3, and the lattice size $N$=987 in (a). Averaging is done over 200 random values of $\phi$.}
\label{scaling.a}
\label{scaling.b}
\par\end{centering}
\end{figure}

\section{KIBBLE-ZUREK MECHANISM}
\label{sec:3}
When $V$ is gradually decreased to approach the critical point, correlation length will diverge as:\\
\begin{equation}
\xi\approx\epsilon^{-\nu}, \epsilon=V-V_{c},
\label{scaling_length}
\end{equation}
where $\epsilon$ is the distance from the critical point and $\nu=0.997\pm0.006$ is a correlation-length exponent extracted from Fig.~\ref{scaling.a}(a). We set $\Delta=0.5$ throughout the paper without loss of generality, and the second-order phase transition occurs at $V_{c}=3$.

The dynamical exponent $z$ can be determined by the scaling of system size $N$ and the relevant gap, i.e.,  $\Delta_{r}=\varepsilon_{0}+\varepsilon_{1}$, which is the sum of energies of the two positive lowest energy quasiparticles \cite{RefDX1,RefDX2,RefDX3}
\begin{equation}
\Delta_{r}\sim N^{-z}.
\label{scaling_gap}
\end{equation}
We use the linear fit to log-log plot of Fig.~\ref{scaling.b}(b) which yields $z=1.373\pm0.023$. The dynamical exponents $z\nu$ determine how the gap vanishes with the distance from the critical point. These critical exponents can be obtained from the study of the the fidelity susceptibility \cite{RefDW} and scaling analysis of superfluid fraction for different lattice sizes\cite{RefDX}. The whole results are also true for other points on the second-order phase transition line, except for the limited conditions of $\Delta=0, -1$. When $\Delta$=0, the Aubry-Andr\'{e} model with $p$-wave superconductivity will return to the Aubry-Andr\'{e} model\cite{RefBH1}. When $\Delta=-1$, the model will return to quasiperiodic Ising model\cite{fermi3,RefF1,RefF2}.

The initial state is deeply prepared in the localized state, and the potential $V$ is slowly changed across the critical point between the critical and the localized phase.

Near the critical point, $\epsilon$ can be approximated by a linear quench:
\begin{equation}
\epsilon\approx-\frac{t}{\tau_{Q}},
\end{equation}
here $\tau_{Q}$ is the quench time. When the state is far away from the critical point, the state is adiabatically evolving. Then, the state crosses the adiabatic region to the diabatic region at a time point when its reaction time $\tau\sim\frac{1}{\Delta_{r}(t)}\sim|\frac{t}{\tau_{Q}}|^{-z\nu}$ equals the time scale $|{\epsilon}/\dot{\epsilon}|$. Thus there exists an intersection in which two timescales are equal, $t=\pm\hat{t}$, where
\begin{figure}[]
\centering
%\vspace{-12pt}
\includegraphics[width=0.48\textwidth]{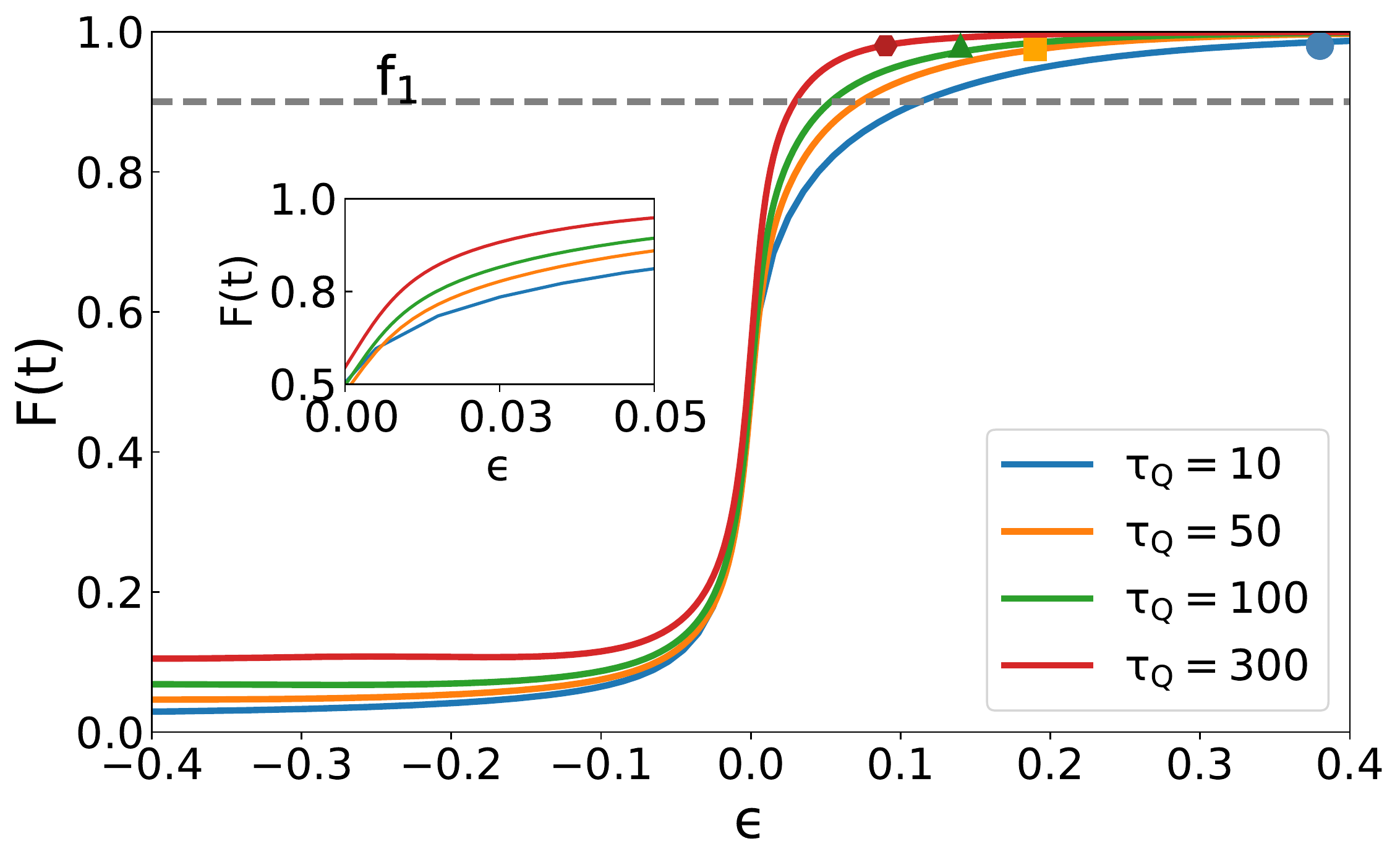}
%\vspace{-12pt}
\caption{The fidelity as a function of $\epsilon$ for four different $\tau_{Q}=10, 50, 100, 300$, and the straight line $f_{1}=0.9$. The blue circle, orange square, green triangle and red hexagon represent four corresponding ``frozen out'' instants. Inset: Enlarged view of $\epsilon$ between 0 and 0.05. The parameters are chosen as $\Delta=0.5$, $V_{c}=3$ and $N=987$.}
\label{scaling.c}
\end{figure}
\begin{equation}
\label{E7}
\hat{t}\sim\tau_{Q}^{z\nu/(1+z\nu)}.
\end{equation}
The time-dependent state is still at the ground state until $t=-\hat{t}$ and $\hat{\epsilon}=\frac{\hat{t}}{\tau_{Q}}\sim\tau_{Q}^{-1/(1+z\nu)}$, with localization length
\begin{equation}
\label{E8}
\hat{\xi}\sim\hat{\epsilon}^{-\nu}\sim\tau_{Q}^{\frac{\nu}{1+z\nu}}.
\end{equation}
In zero-order approximation, the two time points $\pm\hat{t}$ divide the whole evolution into three regimes. Initially, when $t<-\hat{t}$, the state can adjust to the change of the Hamiltonian. However, at $t=-\hat{t}$ this tracking will cease, and the wave-packet does not follow the instantaneous ground state until $\hat{t}$ with a finite localization length $\hat{\xi}$. Afterwards, it is the initial state for the adiabatic process that begins at $\hat{t}$ which is similar to the one ``frozen out'' at $-\hat{t}$.

We should remember that such a ``frozen out'' instant is only a feasible hypothesis. However, it is very helpful to deduce the scaling law. Actually, a realistic system does not exist a sudden change at a certain moment during the evolution, which is a process from the adiabatic to the diabatic regime. Therefore, we can numerically test the KZM hypothesis by solving the critical dynamics, and estimate the frozen instant when the adiabaticity breaking. In this connection, although there is no unique way to quantify adiabatic loss, we use the fidelity $F(t)$,
\begin{equation}
F(t)=|\braket{\psi(t)|\Psi_{0}(t)}|,
\end{equation}
to describe the loss of adiabaticity, which provides a good approximation\cite{RefJH}. Here, $\bra{\psi(t)}$ is the time-evolved state, and $\ket{\Psi_{0}(t)}$ is the instantaneous ground state. In Fig.~\ref{scaling.c}, we plot the time-dependent fidelity $F(t)$ as a function of $\epsilon$ for four different quench rates, and the fidelity $F(t)$ decreases dramatically at the critical point~\cite{RefJI.} From this, we can get the estimated values of the ``frozen out'' instants. The blue circle, orange square, green triangle, red hexagon represent the instants with different $\tau_{Q}$. It is clearly shown that the corresponding ``frozen out'' instants is closer to the critical point as $\tau_{Q}$ increasing. We choose one value represented by the straight line $f_{1}=0.9$, and we can see the fidelity in the four different instants are very close to 1 and away from 0.9. So until the instants, the loss of the adiabaticity is almost zero. But after that, the fidelity tends to fall faster, as shown in Fig.~\ref{scaling.c}.

\begin{figure}[]
\centering
\includegraphics[width=0.48\textwidth]{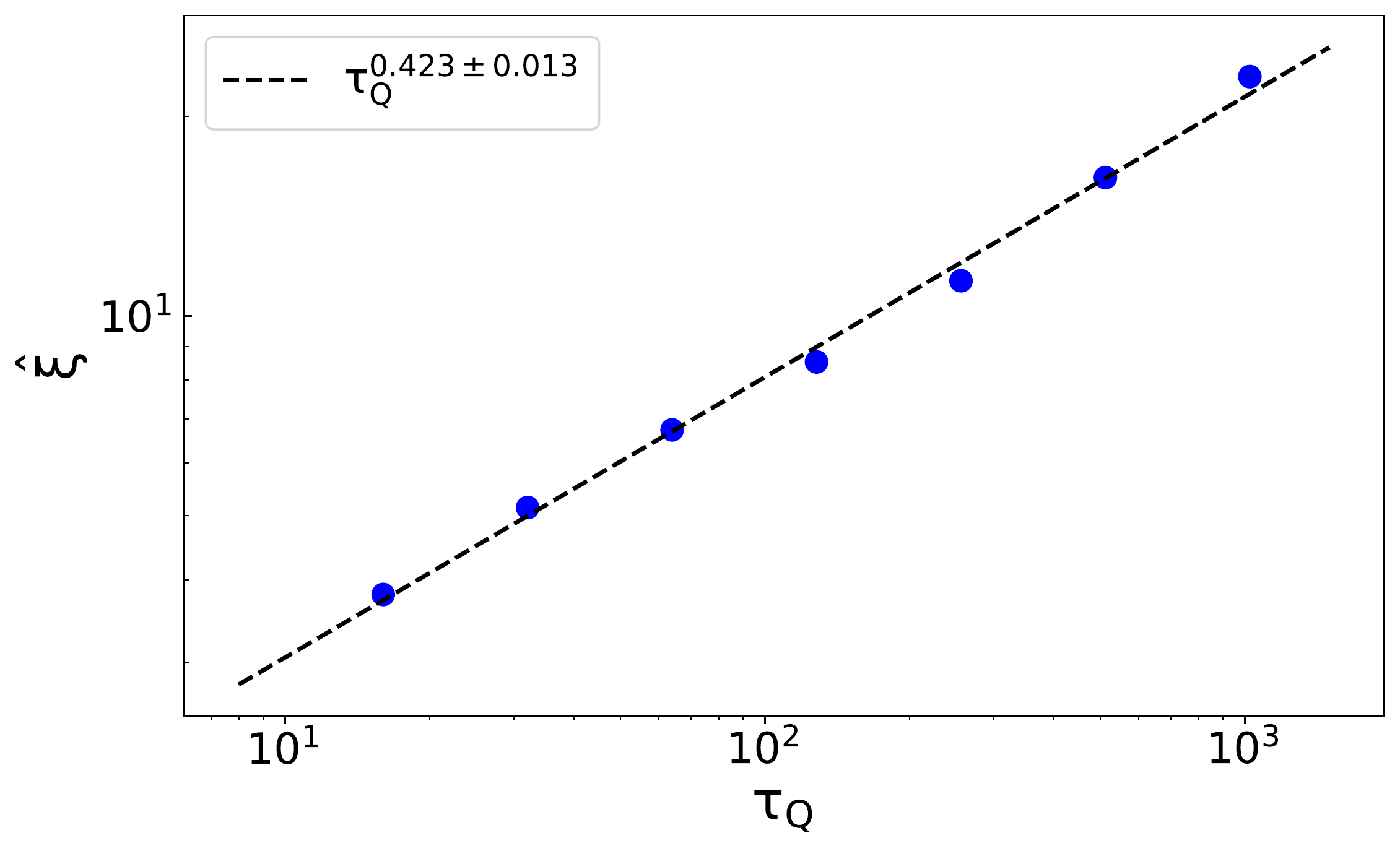}
%\caption{fig1}
\caption{The width of the wave packet as a function of the quench time $\tau_{Q}$ at the critical point. The fitted straight line gives $\hat{\xi}=\tau_{Q}^{0.423\pm0.013}$; cf. Eq.~(\ref{E8}). The parameters are chosen as N=987, $\Delta=0.5$, $V=3$, and $\phi=0$.}
\label{power law.a}
\end{figure}

\subsection{KZ POWER LAWS}
In order to test the KZ scaling, we use smooth tanh-profile $\epsilon(t)=-\tanh{(t/\tau_{Q})}$ starting from $-5\tau_{Q}$ for the sake of suppressing excitation derived from the initial discontinuity of the time derivative $\dot{\epsilon}$ at $-5\tau_{Q}$.

When the system's evolution crosses the adiabatic area at $-\hat{t}$, then in the diabatic area, the localization length $\hat{\xi}$ does not change under the zero-order approximation until the time at $\hat{t}$. In Fig.~\ref{power law.a}, we plot $\hat{\xi}$ estimated by the dispersion of the probability distribution as a function of $\tau_{Q}$ at the critical point $\epsilon=0$. The power law fitting implies $z=1.364$ for $\nu=1$. And $z=1.361$ for $\nu=0.997$ estimated in Fig.~\ref{scaling.a}(a).

The dynamical exponent $z=1.361$ extracted from $\hat{\xi}$ in Fig.~\ref{power law.a} and $z\simeq1.373$ from Fig.~\ref{scaling.b}(b) differ by $1\%$. Similarly, the critical exponent $\nu\simeq0.997$ is also $0.3\%$ away from the value $\nu=1$. The difference is almost the same as the system error. Therefore, within a small error range, our numerical results are consistent with the predicted results.

\begin{figure}
\begin{centering}
\includegraphics[width=1.6in]{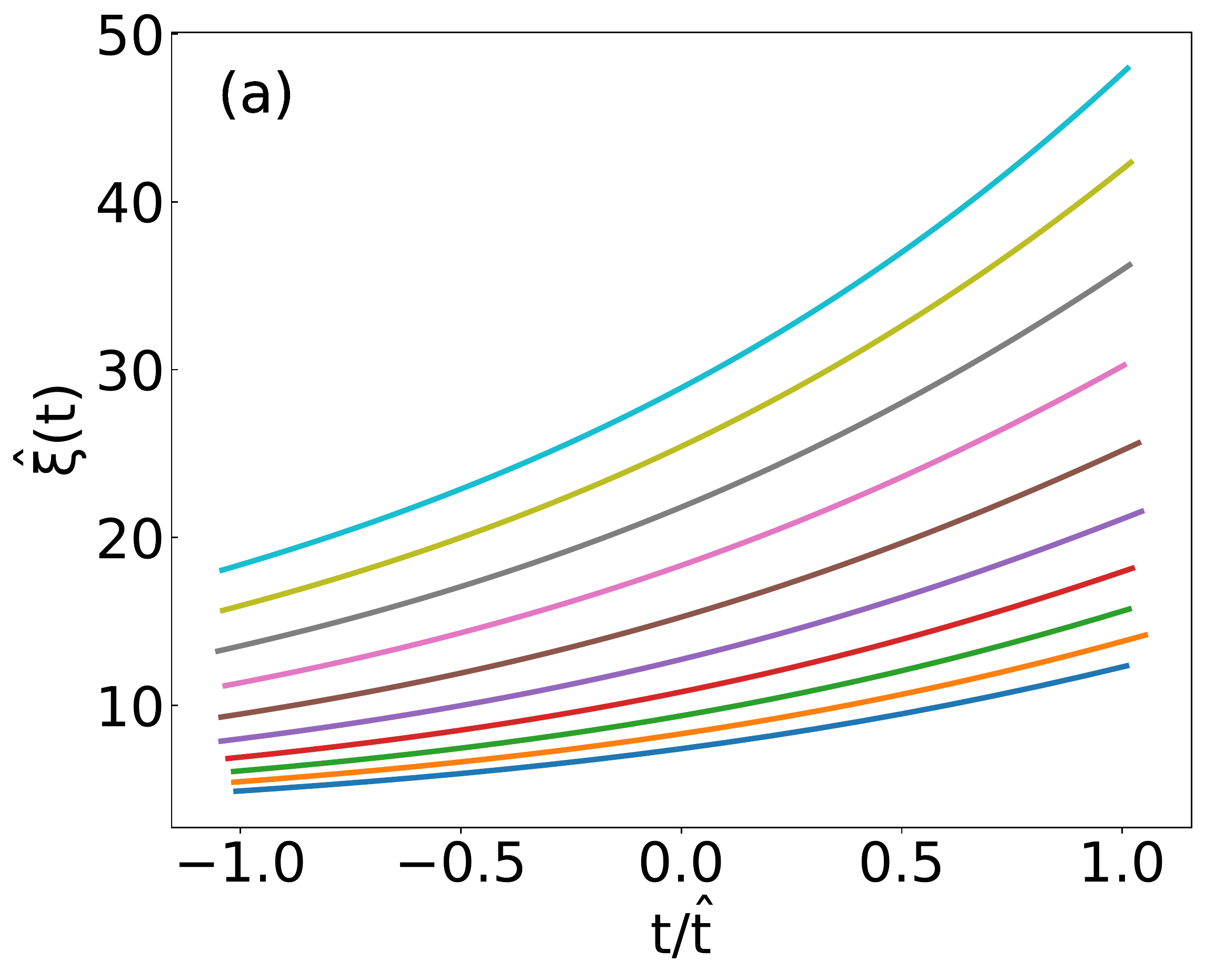}\label{hypothesis.a}\includegraphics[width=1.85in]{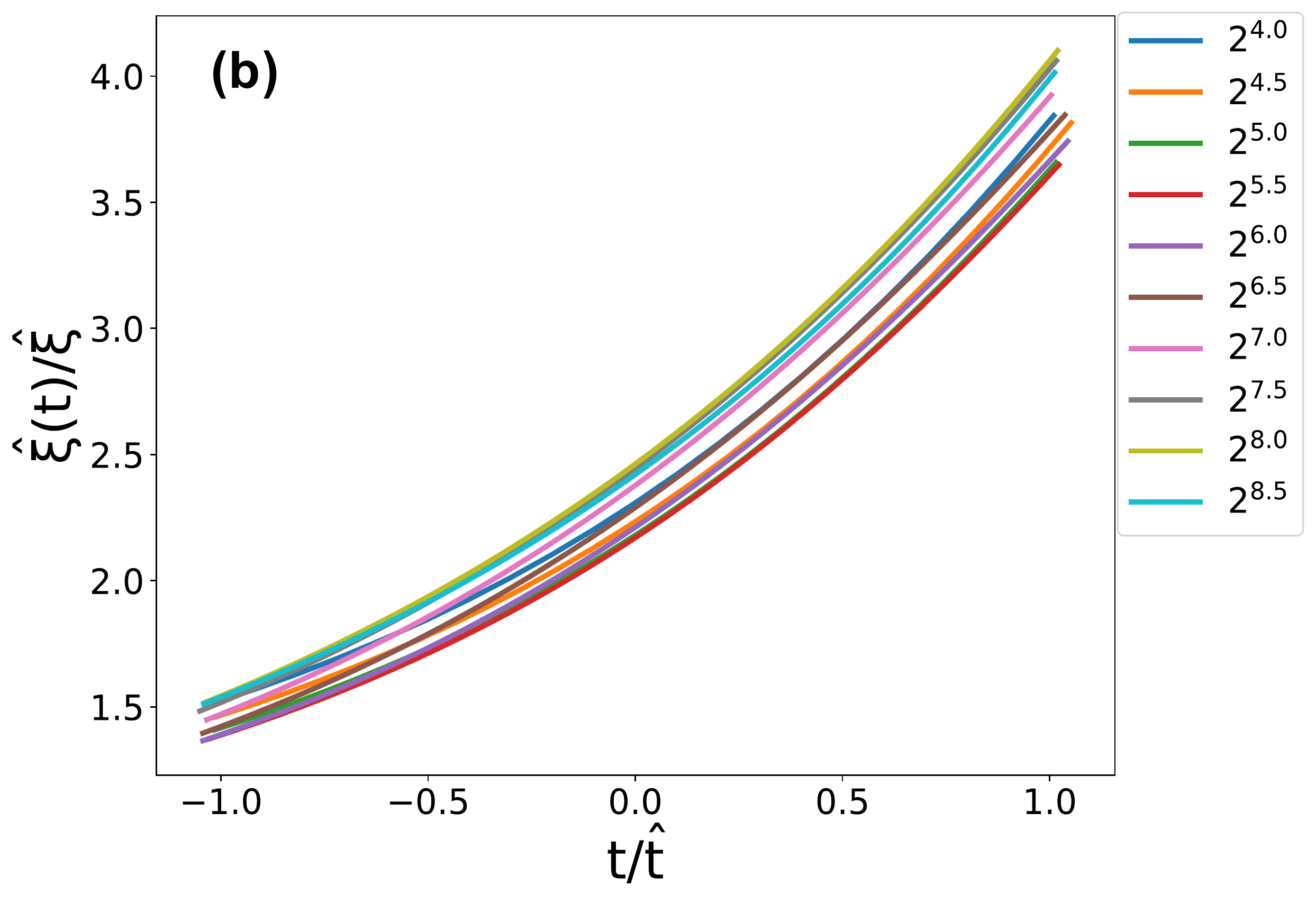}\label{hypothesis.b}

\caption{In (a), the width of the wave packet $\hat{\xi}(t)$ as a function of the scaled time $-\hat{t}$ and $\hat{t}$ which represents the impulse regime. In (b), the scaled width of the wave packet $\hat{\xi}(t)/\hat{\xi}$ and scaled time all collapse to their respective scaling function. The parameters are chosen as Fig.~\ref{power law.a}.}
\label{hypothesis}

\par\end{centering}
\end{figure}

In the impulse area, $\hat{\xi}$ is the relevant scale of length. When $\tau_{Q}\rightarrow\infty$, the adiabatic limited is recovered. $\hat{\xi}$ diverges in the limit and becomes the only relevant scales in the long-wavelength regime. This logic proves the KZ scaling hypothesis \cite{RefC,RefCA,RefCB} for a correlation length $\hat{\xi}(t)$ in the diabatic regime:
\begin{equation}
\hat{\xi}(t)=\hat{\xi}F_{\xi}(t/\hat{t}),
\end{equation}
where $F_{\xi}$ is not a universal function as shown in Fig.~\ref{hypothesis}.

\section{Loschmidt echo}
\label{sec:4}
In the following section, we discuss another nonequilibrium dynamics by suddenly quenching the on-site potential $V$, not only between the localization phase and critical phase separated by the second-order phase transition line, but also between the critical phase and extended phase.

By preparing the initial state as the eigenstate of the Hamiltonian $H(V_{i})$, and then suddenly quenching the Hamiltonian to $H(V_{f})$, we calculate the return probability (Loschmidt echo)\cite{RefF}:
\begin{equation}
L(t,V_{i},V_{f})=\mid G(t,V_{i},V_{f})\mid^2,
\end{equation}
where $G(t,V_{i},V_{f})$ is the return amplitude (a type of Loschmidt echo amplitude):
\begin{equation}
\label{E10}
G(t,V_{i},V_{f})=\braket{\psi(V_{i})|e^{-itH(V_{f})}|\psi(V_{i})},
\end{equation}
where $\psi(V_{i})$ is the eigenstate of the initial Hamiltonian $H(V_{i})$, and $V_{i}$ $(V_{f})$ represents the strength of the initial (final) incommensurate potential. The initial state is chosen to be the ground state of the initial Hamiltonian, and the results are also true for all the other eigenstates.

Then, we illustrate whether the zero points of the Loschmidt echo can be regarded as the signature of the phase transition among the  localized phase, critical phase, and extended phase. To give a more intuitive explanation, we should consider two limiting cases. For these two cases, the initial value of $V_{i}$ is set to 0$(\infty)$ and $V_{f}=\infty(0)$ which can be calculated analytically, whereas the other cases are studied by the numerical methods.

\begin{figure}
\begin{centering}
\includegraphics[width=0.5\columnwidth]{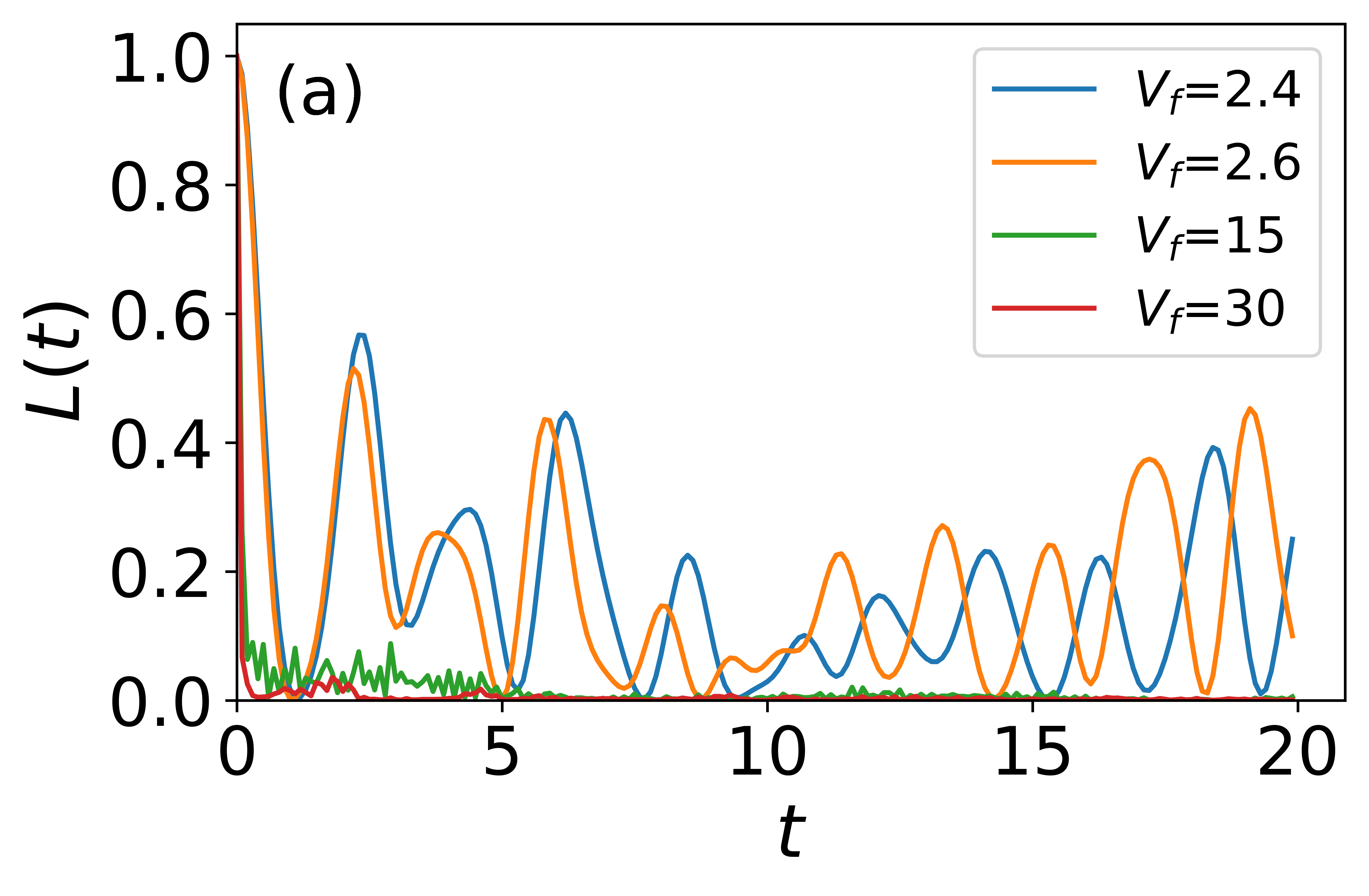}\includegraphics[width=0.5\columnwidth]{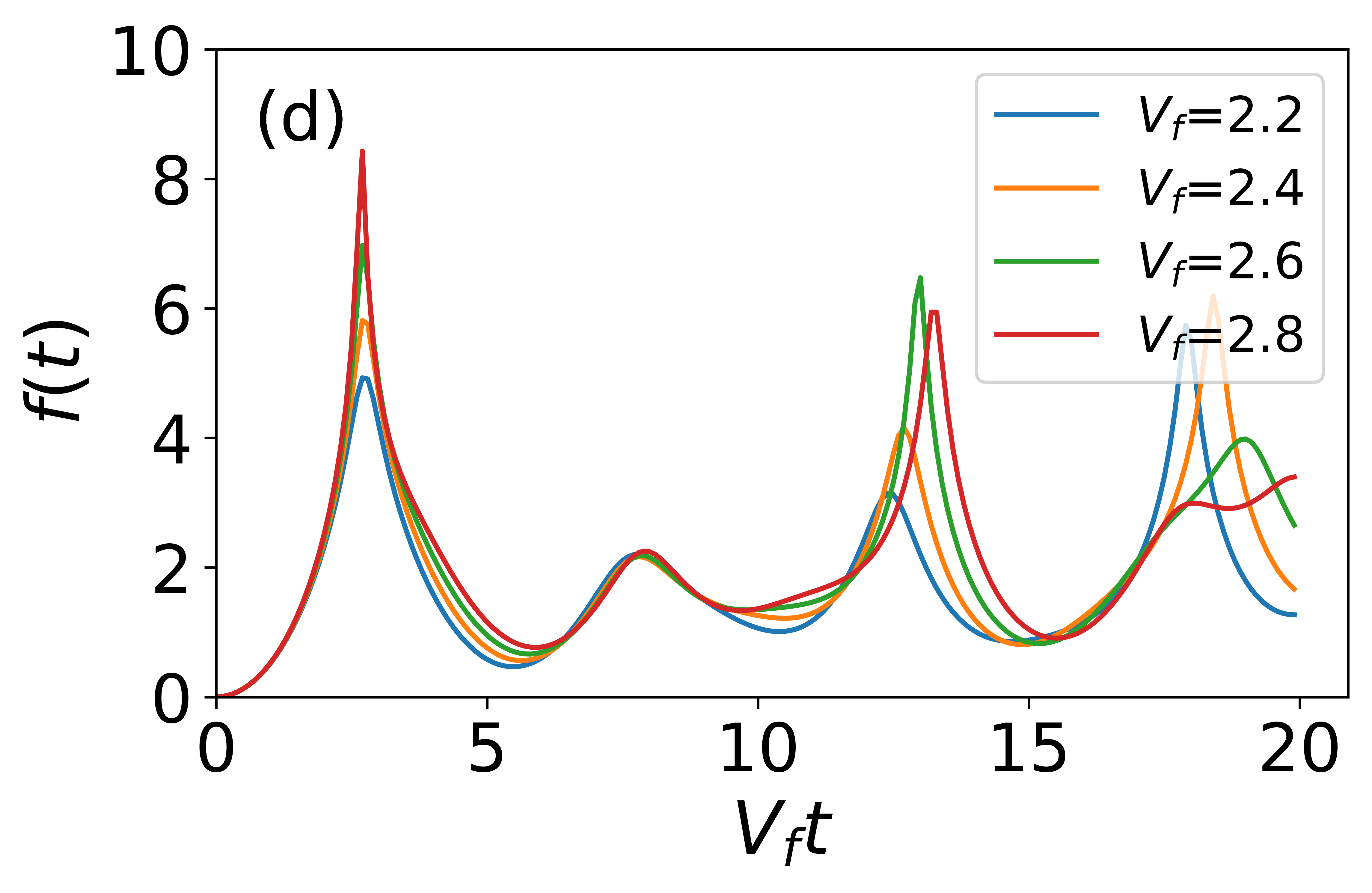}
\par\end{centering}
\begin{centering}
\includegraphics[width=0.5\columnwidth]{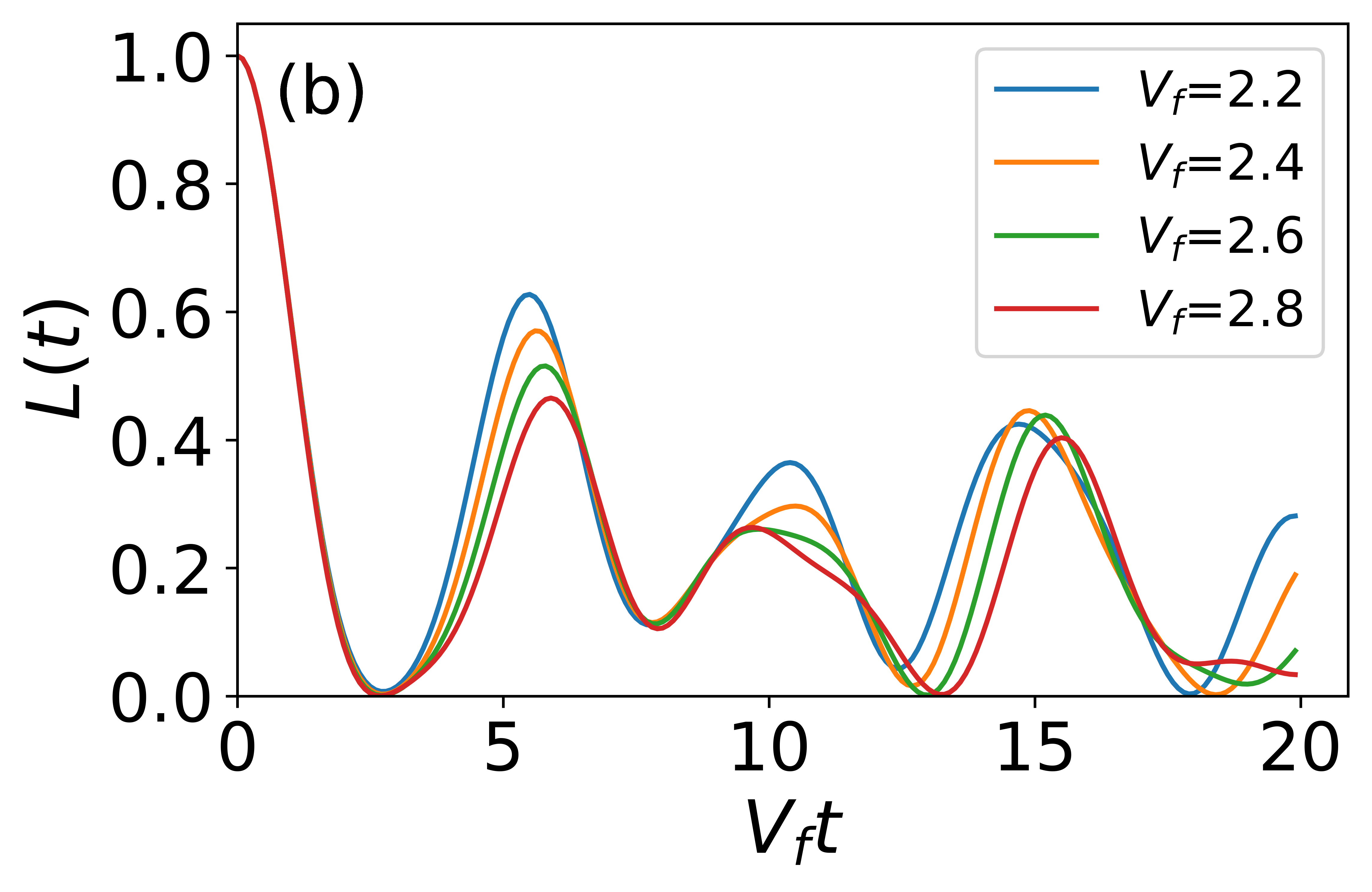}\includegraphics[width=0.5\columnwidth]{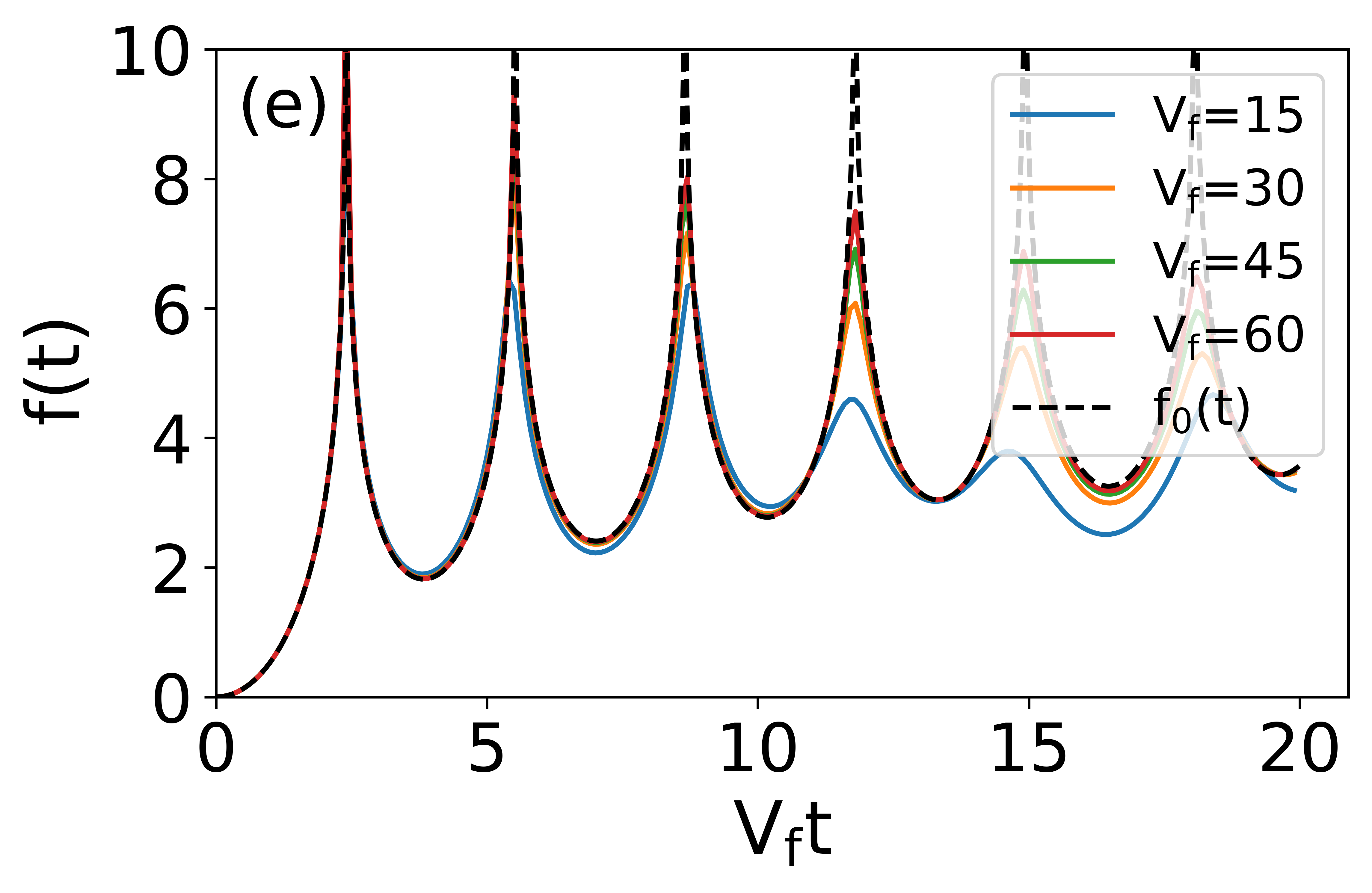}
\par\end{centering}
\begin{centering}
\includegraphics[width=0.5\columnwidth]{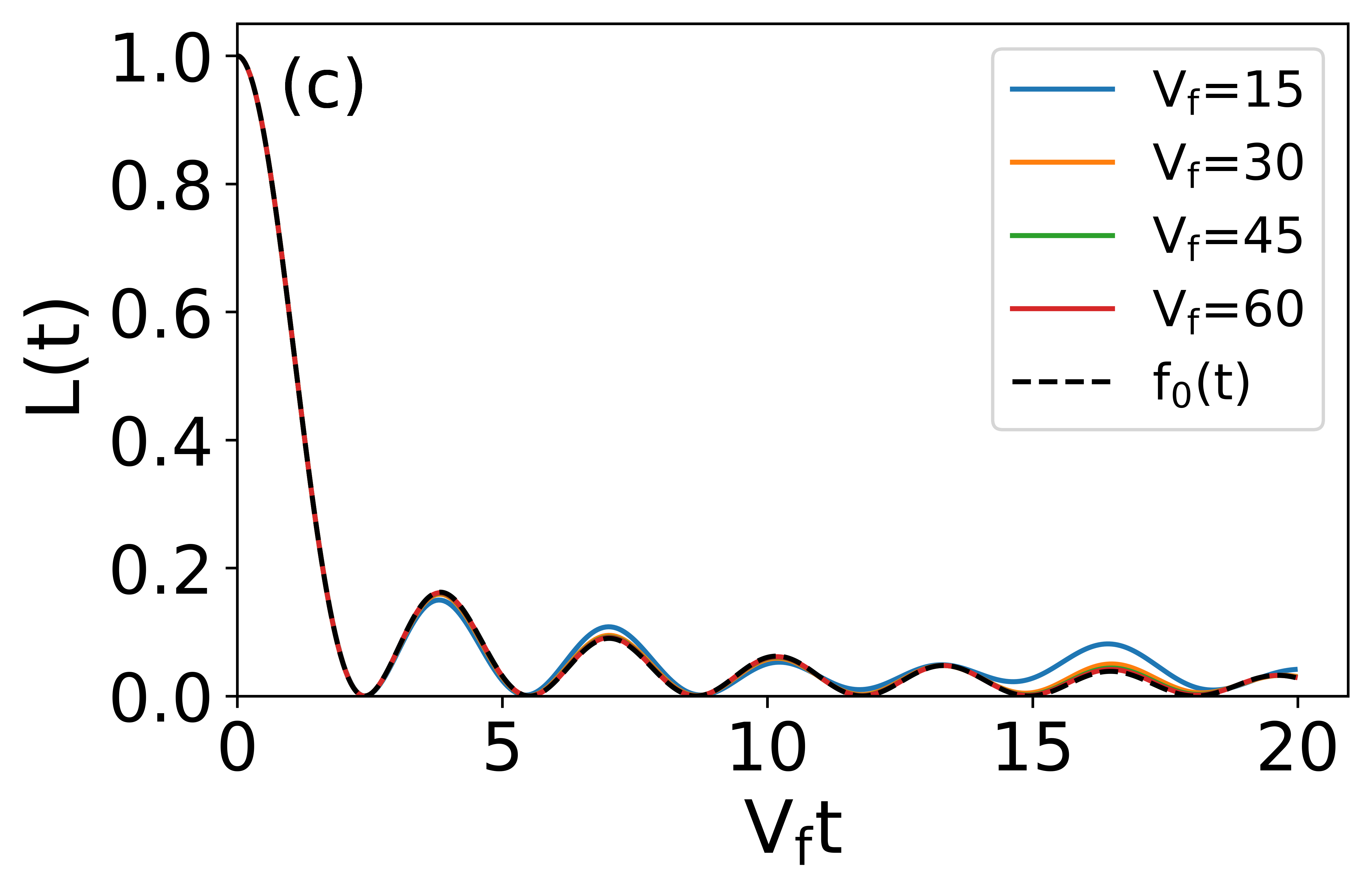}\includegraphics[width=0.5\columnwidth]{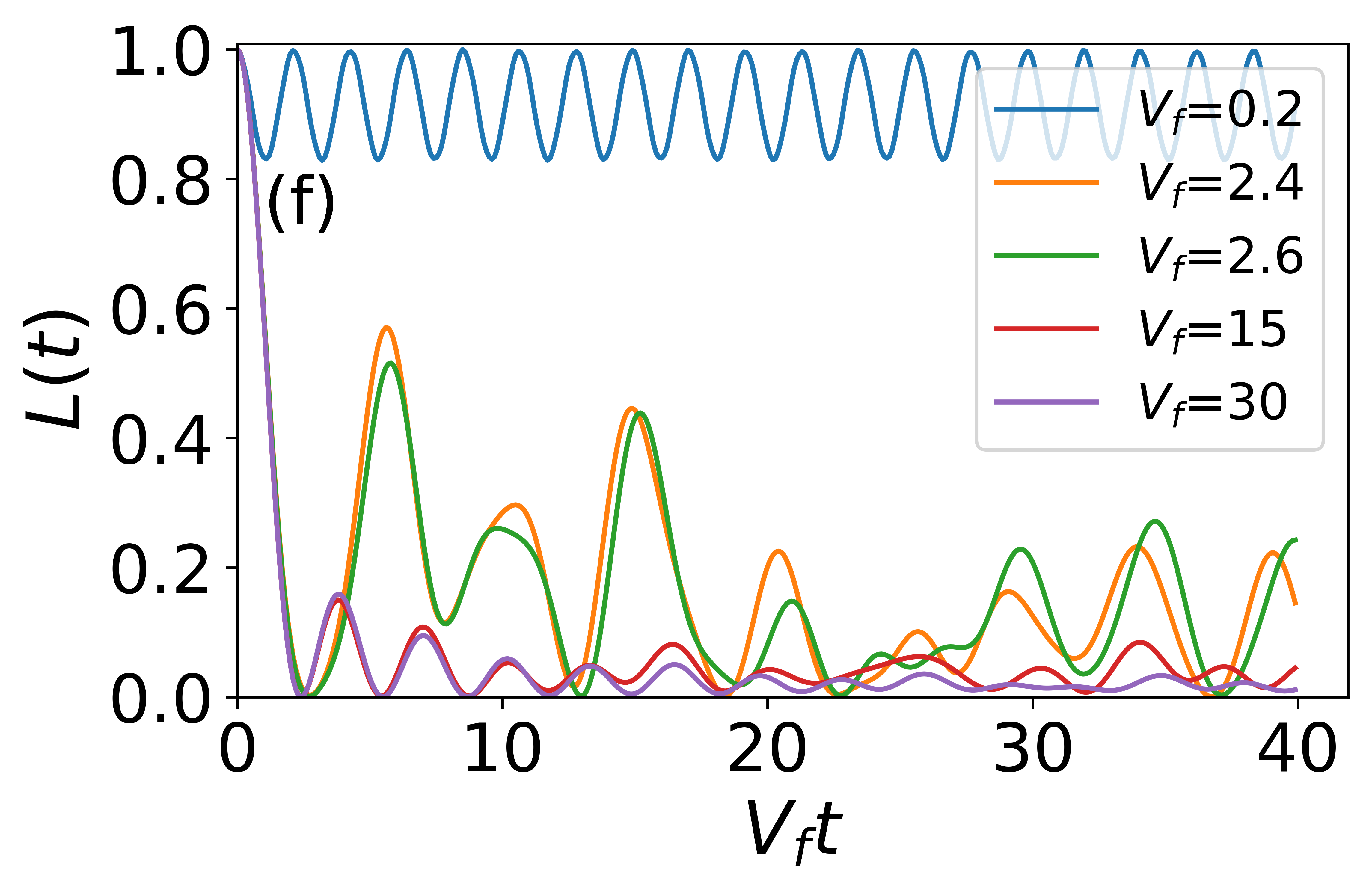}
\par\end{centering}
\caption{The evolution of Loschmidt echo with different $t$ or $V_{f}t$. The system size $N=987$ and the SC paring $\Delta=0.5$. The initial state is set to be the ground state of the Hamiltonian with $V_{i}=0$ . (a) Loschmidt echo versus $t$. (b) and (c) $L(t)$ versus different rescaled time $V_{f}t$. (d) and (e) ``dynamic free energy'' $f(t)$ versus $V_{f}t$. $f_{0}(t)=-log|J_{0}(V_{f}t)|^{2}$ is depicted by the black dotted line. (f) The evolution of the Loschmidt echo for various $V_{f}$ including the extended, critical, localized phases. The Loschmidt echo approaches zero at some different time points. And it has different frequencies for different phases.}
\label{Los-Ex}
\end{figure}

If $V_{i}=0$, the eigenvalues of the Hamiltonian is $\varepsilon_{n}=V_{f}\cos{(2\pi\alpha n)}$, and the corresponding eigenstates are plane wave states $\ket{\phi_{k}(V_{i}=0)}=\frac{e^{-i\pi/4}}{\sqrt{N}}\sum_{j=1}^{N}e^{ikj}c^{\dagger}_{j}\ket{0}$. If $V_{f}=\infty$, the system is in the localized phase, the eigenstates of the Hamiltonian is the localized states $\ket{\Psi_{n}(V_{f}=\infty)}=\sum_{j=1}^{N}{\delta_{jn}c_{j}^{\dagger}\ket{0}}$ with the eigenvalues $\varepsilon_{n}=V_{f}\cos(2\pi \alpha n)$. Then substituting the above results into Eq.~(\ref{E10}), we can get the analytical solution $G_{k}= J_{0}\left(V_{f} t\right)$ [see Appendix~\ref{appendix}], where $J_{0}(V_{f}t)$ is the zero-order Bessel function. It has a number of zeros $x_{n}$ with $n=1, 2, 3, ...$. These zeros mean that the Loschmidt amplitude and the echo can reach zeros at times:
\begin{equation}
t^{*}_{n}=\frac{x_{n}}{V_{f}}.
\label{first_criti_t}
\end{equation}
According to the DQPT theory, the appearance of the zero points in Loschmidt echo can be regarded as the characteristics of the DQPT and it is related to the divergence of the boundary partition function. Because the transition time $ t _ {n}^{*} $ is inversely proportional to $ V_ {f} $, the Loschmidt echo oscillates faster with the increasing $ V_ {f} $ (see Fig.~\ref{Los-Ex}(a)). Then, if we rescale the time $t$ to $V_{f}t$, as shown in Fig.~\ref{Los-Ex}(b)-\ref{Los-Ex}(f), the evolution of the Loschmidt echo shows similar behaviors for the quenching process of different $V_{f}$ as shown in Fig~.\ref{Los-Ex}(b)-\ref{Los-Ex}(c). The initial strength $V_{i}$ is set to $0$ and the SC paring $\Delta=0.5$. It is apparent that the Loschmidt echo for $V_{f}=2.2, 2.4, 2.6, 2.8$ in the critical phase or $V_{f}=15,30,45,60$ in the localized phase oscillates with different frequencies. However, they are all quite similar after rescaling the time $t$ to $V_{f}t$. Except for the smaller $V_{f}=15$, in the localized phase, the numerical results almost coincide with the analytical solution, shown in Fig.~\ref{Los-Ex}(c). Therefore, although the analytical solution is under the condition of $V_{f}\rightarrow \infty$, the above results hold true for large enough $V_{f}$, as shown in Fig.~\ref{Los-Ex}. To see the zero point in Loschmidt echo more clearly, we calculate the ``dynamical free energy", defined as $f(t)=-log|G(t)|^{2}$. $f(t)$ will be divergent at the time point $t=t_{n}^{*}$\cite{RefDA,RefDB}. In Fig.~\ref{Los-Ex}(d) and Fig.~\ref{Los-Ex}(e), $f(t)$ is plotted as a function of different $V_{f}t$ with $V_{f}$ in the localized phase or in the critical phase. Obviously it reaches the peaks at the critical times $t_{n}^{*}$, especially when $V_{f}$ gets closer to $\infty$.

In Fig.~\ref{Los-Ex}(f), we calculate $L(t)$ as a function of the scaled time $V_{f}t$ with a series of final value taken in different phases. When $V_{f}<2|J-\Delta|$, the Loschmidt echo can not reach the zero even for long time evolution, because $V_{i}$ and $V_{f}$ are in the same phase. However, when the final value $V_{f}$ is in the critical phase or in the localized phase, $L(t)$ shows similar oscillations with different $\Delta_{f}$ and reaches zeros. And the time interval of the Loschmidt echo has different in approaching the zero points between the critical and localized phase. By noticing that the condition of $V_{f}\rightarrow \infty$ can not be met in the critical phase, the analytical result of $J_{0}(V_{f}t)$ is no longer applicable.

\begin{figure}
\begin{centering}
\includegraphics[width=0.5\columnwidth]{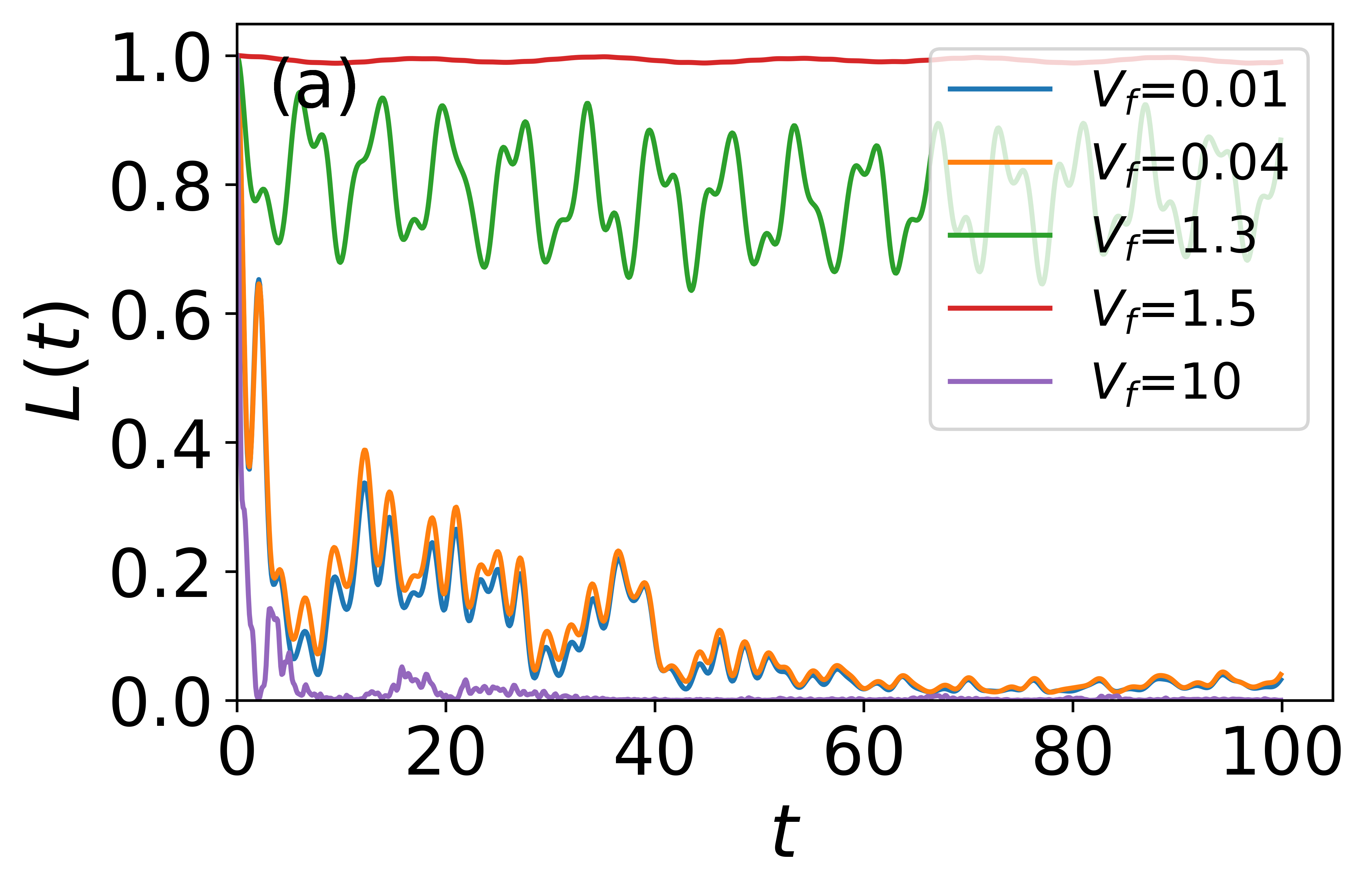}\includegraphics[width=0.5\columnwidth]{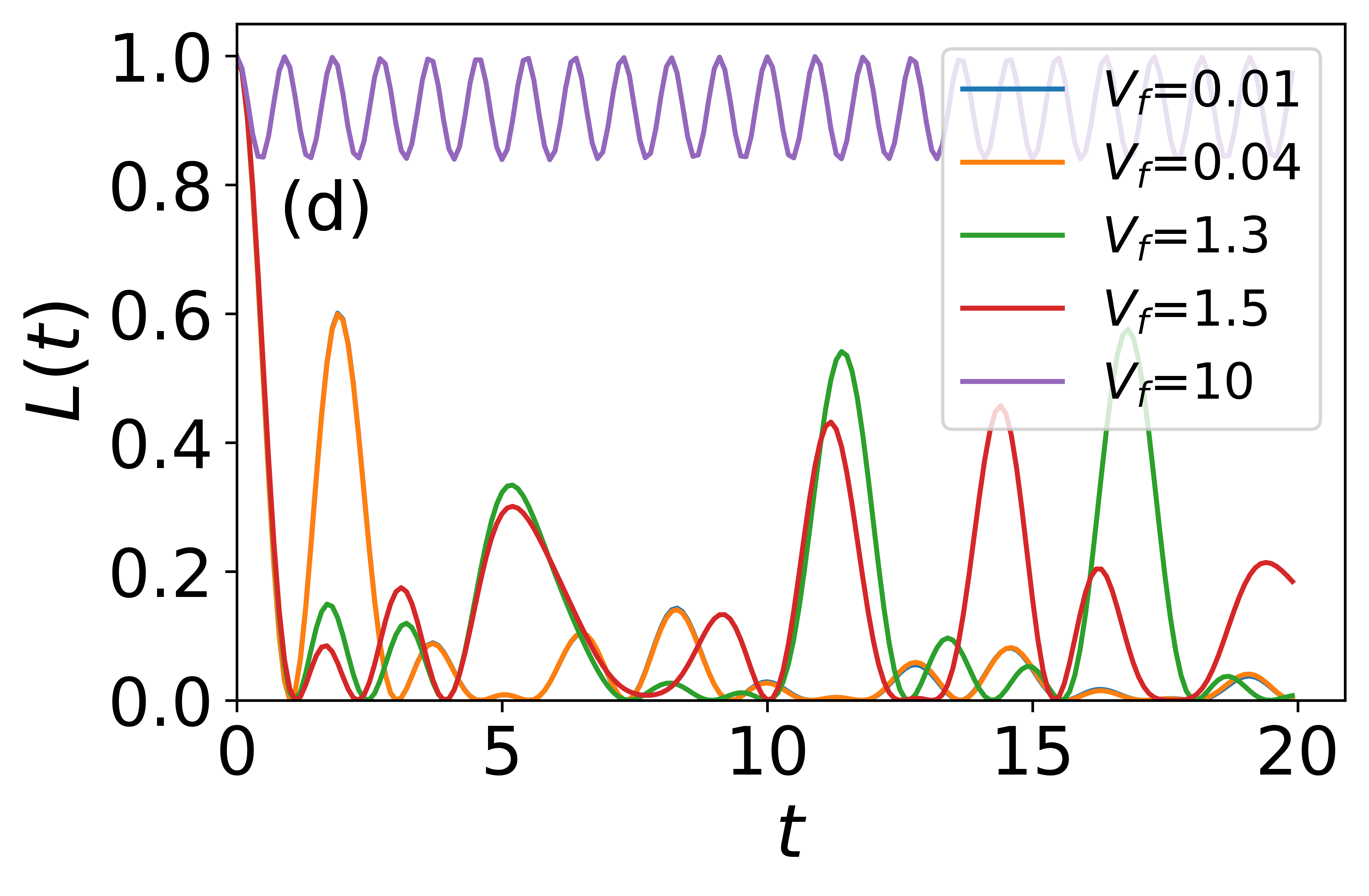}
\par\end{centering}
\begin{centering}
\includegraphics[width=0.5\columnwidth]{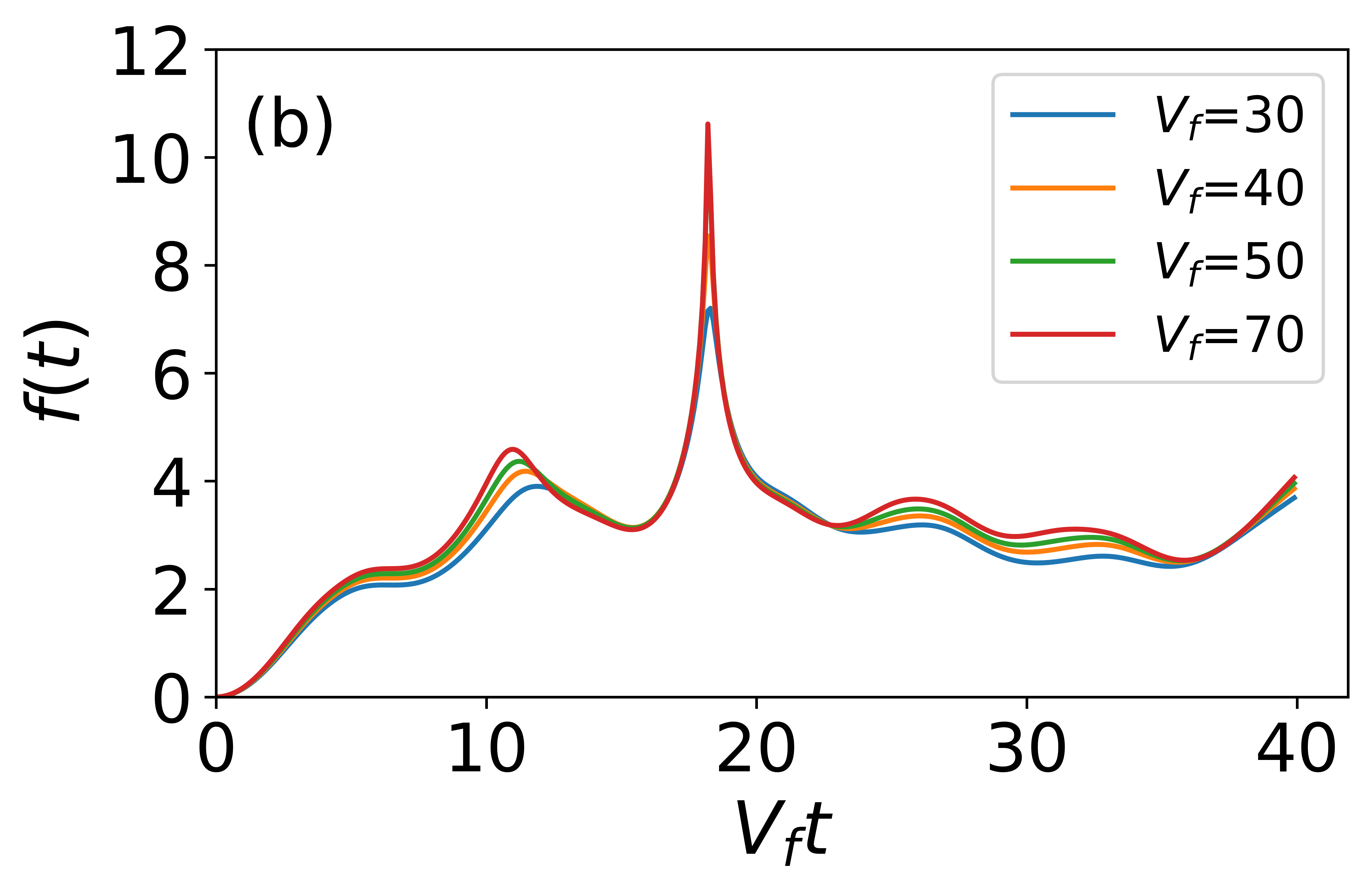}\includegraphics[width=0.5\columnwidth]{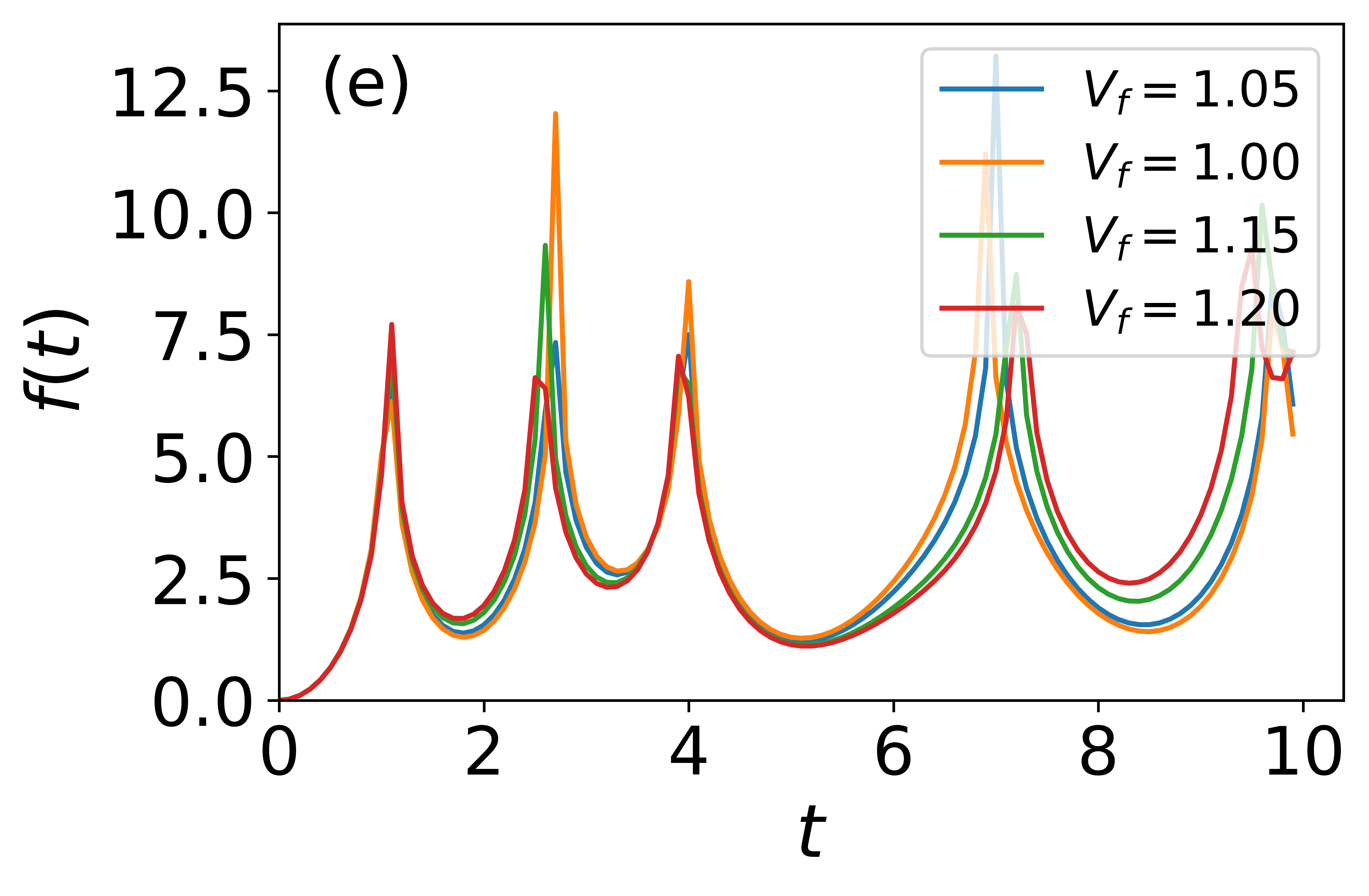}
\par\end{centering}
\begin{centering}
\includegraphics[width=0.5\columnwidth]{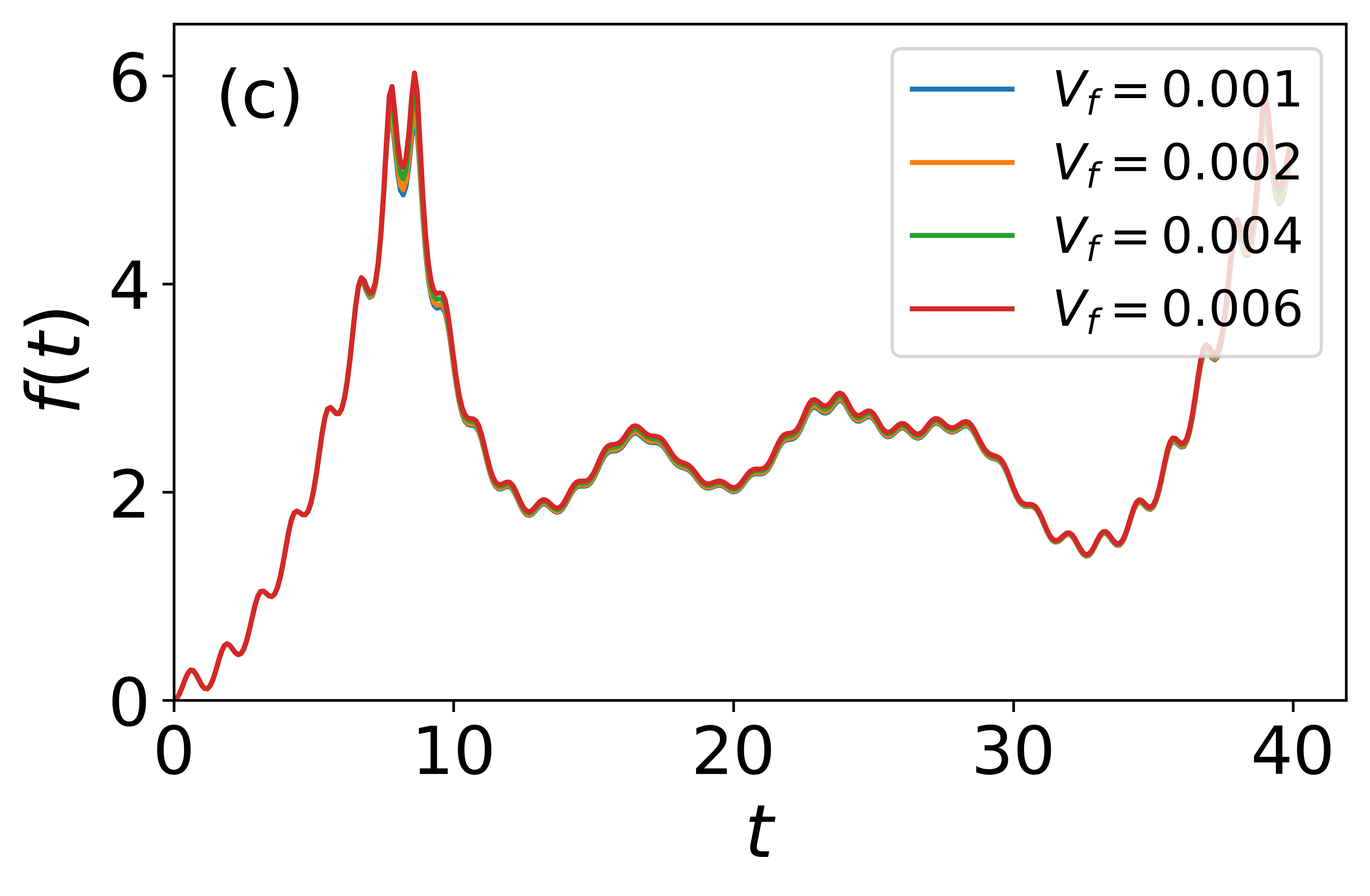}\includegraphics[width=0.5\columnwidth]{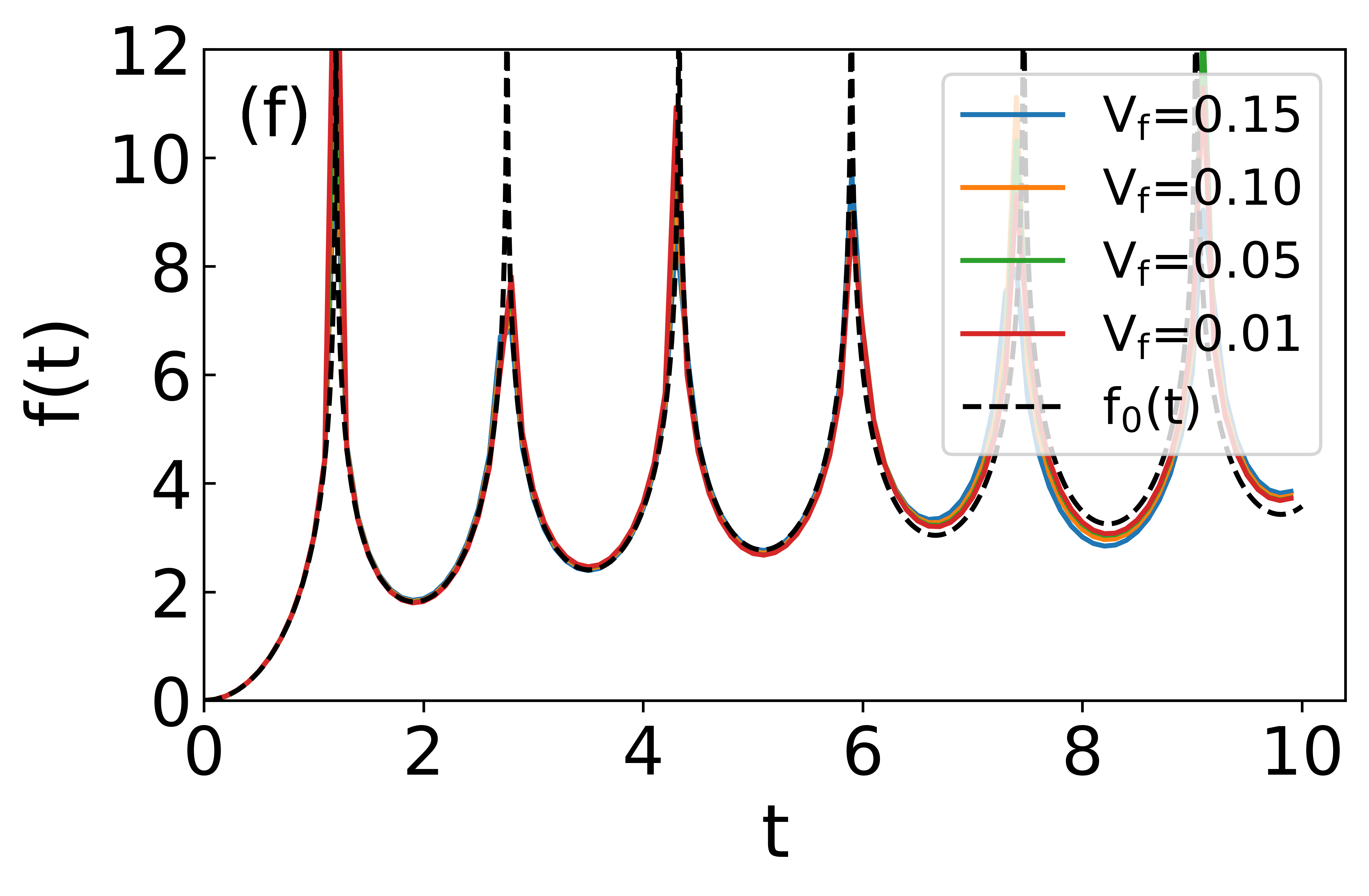}
\par\end{centering}

\caption{The evolution of Loschmidt echo with different $V_{f}t$ and $t$. The system size $N=987$, the SC paring $\Delta=0.5$, and the initial state is set to the ground eigenstate of the Hamiltonian of $V_{i}=2.6$ for (a), (b), and (c), but of $V_{i}=100$ $(d)$, $(e)$, and $(f)$. The initial state is the ground state of the Hamiltonian. (a) Loschmidt echo versus $t$; (b-c) $f(t)$ versus different rescaled time $V_{f}t$ and time $t$; (d) The evolution of the Loschmidt echo for various $t$. (e-f) ``dynamic free energy''  $f(t)$ versus $t$. The black dotted line corresponds to $f_{0}(t)=-log |J_{0}(2Jt)|^{2}$ and the SC paring $\Delta=0.05$. In Figs (d), (e) and (f), the initial states are all in the localized phase. When $V_{f}<1$, the Loschmidt echo will approach zero at the same time interval which is independent of $V_{f}$.}
\label{Los-lc}
\end{figure}

Furthermore, we study the quenching process from a strong disorder strength $V_{i}$ to $V_{f}=0$. The system is initially prepared in the eigenstate of the localized phase, then is quenched into the extended regime. Similar to the above analysis [see Appendix~\ref{appendix}], we get the Loschmidt amplitude $G_{n}=J_{0}(2Jt)$, and the zero points of the Loschmidt echo appear at times:
\begin{equation}
t_{n}^{*}=\frac{x_{n}}{2J},
\end{equation}
which is inversely proportional to the hopping amplitude $2J$, different from Eq.~(\ref{first_criti_t}). The transition time $t_{n}^{*}$ is independent of $V_{f}$ which means that for the different $V_{f}$ the dynamical free energy has almost the same behaviors. Moreover, the return amplitude is insensitive to $V_{i}$, as long as $V_{i}$ is large enough, even in the critical phase.

In Fig.~\ref{Los-lc}, the Loschmidt echo and the dynamical free energy $f(t)$ as a function of the rescaled time $V_{f}t$ or time $t$. But different from Fig.~\ref{Los-Ex}, the initial system here is in the critical phase or localized phase. In the left panel of Fig.~\ref{Los-lc}, the initial state is prepared in the critical phase,  and in the right panel of the Fig.~\ref{Los-lc}, the initial state is set in the localized phase. Therefore, it is different from the previous analytical result. When $V_{i}=2.6$ and $V_{f}\gg3$, we rescale $t$ to $V_{f}t$. However, the rescaling is not needed when $V_{f}\ll1$, shown in Fig.~\ref{Los-lc}. In Fig.~\ref{Los-lc}(a), we set $V_{i}=2.6$ with $V_{f}=0.01,0.04, 1.3, 1.5, 10$, as long as $V_{f}<2|J-\Delta|$ or $V_{f}>2|J+\Delta|$ the Loschmidt echo will approach zero immediately, but when $V_{f}=1.3, 1.5$ in the critical phase, $L(t)$ will never approach zero during the time evolution. In Fig.~\ref{Los-lc}(b), $f(t)$ also shows similar behavior for different $V_{f}$ after rescaling the time $t$ to $V_{f}t$, due to the final value of the potential $V_{f}\gg V_{i}$. But the shape of the curve is different from $J_{0}(2Jt)$, because the initial state is in the critical phase. For Fig.~\ref{Los-lc}(c), the same reason leads to mismatch between the peak shape and $J_{0}(2Jt)$. In analog to $V_{i}=2.6$, we set $V_{i}=100$ and take a series of $V_{f}$. We find that the Loschmidt echo approaches zero when $V_{f}<2|J+\Delta|$ and it is also true for $V_{f}<2|J-\Delta|$ in Fig.~\ref{Los-lc}(d)-\ref{Los-lc}(f). From Fig.~\ref{Los-lc}(e), \ref{Los-lc}(f), for $V_{f}$ in the critical phase and extended phase, respectively, $f(t)$ shows similar behaviors. In Fig.~\ref{Los-lc}(f), when the SC paring $\Delta=0.05$, the behaviors of $f(t)$ with different $V_{f}$ almost coincide with the analysis result $J_{0}(2Jt)$. As a result, when $V_{f}$ approaches the limit of $V_{f}=0$, the analytical result $G_{n}=J_{0}(2Jt)$ is a good approximation.

\section{Conclusion}
\label{sec:5}
In summary, we have studied the different nonequilibrium dynamics of the 1D AAH model with $p$-wave superconductivity in two different ways. Firstly, a linear ramp crossing the localization-critical phase transition line is not adiabatic. By linearly fitting for the localization length near the critical point, we obtain the critical exponents $z\nu$ with $\nu=0.997$ which is the same as Aubry-Andr\'{e} model, and the dynamical exponent $z=1.373$ which is different from the one in the literature\cite{RefDW,RefDX,RefBH1}. Except for the point $\Delta=0$, the critical exponents are almost the same for all the second-order phase transition line $V=2|J+\Delta|$. We also tried a series of different quenching directions. The critical exponents are the same as what we obtained. Furthermore, we have analyzed the correlation length also the rescaled correlation length as a function of the quench time at the phase transition point within the impulse regime between $-\hat{t}$ and $\hat{t}$. The results are all consistent with the KZ scaling hypothesis. Our results indicate that KZM dominates the nonadiabatic dynamics of the one-dimensional incommensurate system with the localized-critical phase transition.

Next, by using the Loschmidt echo we study the sudden quench dynamics of the time evolution of the AAH model with $p$-wave SC pairing. The results show that the Loschmidt echo reaches zeros as long as the initial and the final system are not in the same phase, which is also true for the critical phase. Especially, if $V_{i}$ is in the critical phase, $L(t)$ and $f(t)$ show similar behaviors when the change of $V$ has the same direction as the two limit cases mentioned before \cite{RefE}. Our research results indicate that the zeros of the Loschmidt echo manifest the dynamic characteristics in the incommensurate system, including the localized phase, critical phase and extended phase.

Here we want to address some interesting issues to be investigated further. We first observe that the role played by the incommensurability, i.e. the irrational number $\alpha$ on the QP potential, was only slightly explored. It has been known that $\alpha$ determines the universality class and the exotic non-power-law behavior~\cite{alpha1,alpha2,alpha3}.  Finally, it is worth to study how the nonequilibrium dynamics of generalized AAH models, for instance QP modulation of the hopping and on-site potential \cite{RefJ}, follows the KZM scaling hypothesis and displays signatures of DQPTs captured by the Loschmidt echo dynamics.

\begin{acknowledgments}
This work is supported by NSF of China under Grant Nos. 11835011 and 11774316.
\end{acknowledgments}

\appendix
\section{}
\label{appendix}
Firstly, $V_{i}$ is set to 0 and the system is initially prepared in the extended phase with periodic boundary condition, that is, a plane wave state is the eigenstate of the Hamiltonian:
\begin{equation}
\label{E12}
\ket{\phi_{k}(V_{i}=0)}=\frac{e^{-i\pi/4}}{\sqrt{N}}\sum_{j=1}^{N}e^{ikj}c^{\dagger}_{j}\ket{0},
\end{equation}
where the wave vector $k=\frac{2\pi(l-\frac{N}{2})}{aN}\in(-\frac{\pi}{a},\frac{\pi}{a}]~(l=1,\ldots,N)$  in the Brillouin zone.  With the eigenvalues $\varepsilon_{k}$ of the initial Hamiltonian~$H(V_{i})$:
\begin{equation}
\label{E19}
\varepsilon_{k}=2\sqrt{(J\cos{ka})^2+(\Delta \sin{ka})^2}.
\end{equation}
When $V_{f}\rightarrow\infty$, the eigenstates of the Hamiltonian become:
\begin{equation}
\label{E14}
\ket{\Psi_{n}(V_{f}=\infty)}=\sum_{j=1}^{N}{\delta_{jn}c_{j}^{\dagger}\ket{0}}.
\end{equation}
Here, $\ket{\Psi_{n}(V_{f})}$ represents the $n$-th eigenstates of the quenched Hamiltonian. The corresponding eigenvalues $\varepsilon_{n}$ of the quenched Hamiltonian is:
\begin{equation}
\label{E13}
\varepsilon_{n}=V_{f}\cos{(2\pi\alpha n)}.
\end{equation}
For a sudden quench, the system crosses from the initial value $V_{i}$ to final value $V_{f}$. For simplicity, we use $\ket{k}$ replace $\ket{\phi_{k}(V_{i}=0)}$. So substituting Eqs.~(\ref{E12}),~(\ref{E14}) and (\ref{E13}) into Eq.~(\ref{E10}), the return amplitude can rewritten as
\begin{align}
\label{E16}
G_{k}(t) & =\braket{k|e^{-i H(V_{f}) t}|k}\nonumber \\
 & =\sum_{n}\braket{k|e^{-i H(V_{f})t}| \Psi_{n}(V_{f})}\braket{\Psi_{n}(V_{f}) | k}\nonumber \\
 & =\sum_{n} e^{-i \varepsilon_{n} t}|\braket{\Psi_{n}(V_{f}) |k} |^{2}\nonumber \\
 & =\frac{1}{N}\sum_{n=1}{e^{-iV_{f}t\cos{(2\pi\alpha n)}}}.
\end{align}
Because of the irrational number $\alpha$, the phase $2\pi\alpha n$ $(n = 1,\ldots ,N)$ modulus $2\pi$ is sett  randomly between $-\pi$ and $\pi$ when we sum over  from 1 to the large $N$. So we can approximately replace the summation by the integration
\begin{equation}
\label{E17}
G_{k}(t) \approx \frac{1}{2 \pi} \int_{-\pi}^{\pi} e^{-i V_{f} t \cos \theta} d \theta=J_{0}\left(V_{f} t\right),
\end{equation}
where $J_{0}(V_{f}t)$ is the zero-order Bessel function. According to the nature of Bessel function, we know that the zero-order Bessel function $J_{0}(x)$ has a series zero-point $x_{n}$ with $n=1, 2, 3, ...$.
In the first case, the Loschmidt echo will reach zero at times:
\begin{equation}
t_{n}^{*}=\frac{x_{n}}{V_{f}}.
\label{f_critical time}
\end{equation}

Conversely, we consider another limit, the quenching process from a strong disorder strength $V_{i}\rightarrow\infty$ to the final $V_{f}=0$. By substituting Eqs.~(\ref{E12}), ~(\ref{E14}) and Eq.~(\ref{E19}) into Eq.~(\ref{E10}), we can get the return amplitude:
\begin{align}
G_{n}(t) &=\braket{n|e^{-iH(V_{f})t}|n}\notag\\
&=\sum_{k}\braket{n|e^{-2it\sqrt{(J\cos{ka})^2+(\Delta \sin{ka})^2}}|k}\braket{k|n}\notag\\
&=\sum_{k}e^{-2it\sqrt{(J\cos{ka})^2+(\Delta \sin{ka})^2}}|\braket{k|n}|^2\notag\\
&=\frac{1}{N}\sum_{k}e^{-2it\sqrt{(J\cos{ka})^2+(\Delta \sin{ka})^2}},
\end{align}
where $\ket{n}$ denotes $\ket{\Psi_{n}(V_{i}=\infty)}$. When $\Delta\ll J$ and in the large $N$ limit, the sum can be transformed into an integral. The same is true for $\Delta\gg J$
\begin{align}
G_{n}(t) & =\frac{a}{2\pi}\int_{-\frac{\pi}{a}}^{\frac{\pi}{a}}e^{-2iJt\cos ka}dk\notag\\
& =J_{0}(2Jt).
\label{second limit}
\end{align}

Therefore, the Loschmidt echo gets zero at times:
\begin{equation}
t_{n}^{*}=\frac{x_{n}}{2J},
\end{equation}
which are $1/2J$ of the zeros of the zero-order Bessel function $J_{0}(x)$, different to Eq.~(\ref{f_critical time}).

%\nocite{*}
\bibliography{apssamp}% Produces the bibliography via BibTeX.
\end{document}